\documentclass[11pt]{article}
\pdfoutput=1
\usepackage{jcapmod}
\usepackage{lmodern}
\usepackage{booktabs}
\usepackage{bm}
\usepackage[english]{babel}
\usepackage[makeroom]{cancel}
\usepackage[per-mode=reciprocal, mode=math, group-digits=integer]{siunitx}[=v2]
\usepackage{subfig}
\usepackage[babel=true]{microtype}

\numberwithin{equation}{section}
\allowdisplaybreaks[1]

\renewcommand{\d}{\mathrm{d}}
\newcommand{\ee}{\mathrm{e}}

\newcommand{\As}{A_\mathrm{s}}
\newcommand{\ns}{n_\mathrm{s}}
\newcommand{\Neff}{N_\mathrm{eff}}
\newcommand{\Nphase}{N_\mathrm{eff}^{\delta\phi}}
\newcommand{\Nmultipole}{N_\mathrm{eff}^{\delta\ell}}

\definecolor{goldenrod}{RGB}{218, 165, 32}
\definecolor{navy}{RGB}{0, 0, 128}
\definecolor{tabblue}{RGB}{31, 119, 180}
\definecolor{tabcyan}{RGB}{0, 200, 210}
\definecolor{tabgray}{RGB}{100, 100, 100}
\definecolor{tabgreen}{RGB}{44, 160, 44}
\definecolor{tabred}{RGB}{214, 39, 40}
\definecolor{tabpink}{RGB}{255, 141, 161}
\DeclareSIUnit{\parsec}{pc}
\DeclareSIUnit{\Mpc}{\mega\parsec}

\DeclareRobustCommand{\SkipTocEntry}[4]{}

\begin{document}

\pagenumbering{roman}
\begin{titlepage}
	\baselineskip=15.5pt \thispagestyle{empty}
	
	\bigskip\
	
	\vspace{1cm}
	\begin{center}
		{\fontsize{20.74}{24}\selectfont \sffamily \bfseries Free-Streaming Neutrinos and Their Phase Shift\\[8pt]in Current and Future CMB Power Spectra}
	\end{center}
	
	\vspace{0.2cm}
	\begin{center}
		{\fontsize{12}{30}\selectfont Gabriele Montefalcone,$^{\bigstar}$ Benjamin Wallisch,$^{\spadesuit,\blacklozenge,\bigstar}$} and Katherine Freese\hskip1pt$^{\bigstar,\spadesuit,\blacklozenge}$
	\end{center}
	
	\begin{center}
		\vskip8pt
		\textsl{$^\bigstar$ Texas Center for Cosmology and Astroparticle Physics, Weinberg Institute for Theoretical\\Physics, Department of Physics, The University of Texas at Austin, Austin, TX~78712, USA}
		
		\vskip8pt
		\textsl{$^\spadesuit$ Oskar Klein Centre, Department of Physics, Stockholm University, 10691~Stockholm, SE}
		
		\vskip8pt
		\textsl{$^\blacklozenge$ Nordita, KTH Royal Institute of Technology and Stockholm University, 10691~Stockholm, SE}
	\end{center}

	\vspace{1.2cm}
	\hrule \vspace{0.3cm}
	\noindent {\sffamily \bfseries Abstract}\\[0.1cm]
	The cosmic neutrino background and other light relics leave distinct imprints in the cosmic microwave background anisotropies through their gravitational influence. Since neutrinos decoupled from the primordial plasma about one second after the big bang, they have been free-streaming through the universe. This induced a characteristic phase shift in the acoustic peaks as a unique signature. In this work, we constrain the free-streaming nature of these relativistic species and other light relics beyond the Standard Model of particle physics by establishing two complementary template-based approaches to robustly infer the size of this phase shift from the temperature and polarization power spectra. One template shifts the multipoles in these spectra, while the other novel template more fundamentally isolates the phase shift at the level of the underlying photon-baryon perturbations. Applying these methods to Planck data, we detect the neutrino-induced phase shift at about~$10\sigma$~significance, which rises to roughly~$14\sigma$ with additional data from the Atacama Cosmology Telescope and the South Pole Telescope. We also infer that the data is consistent with the Standard Model prediction of three free-streaming neutrinos. In addition, we forecast the capabilities of future experiments which will enable significantly more precise phase-shift measurements, with the Simons Observatory and~\mbox{CMB-S4} reducing the $1\sigma$~uncertainties to roughly~4.3\% and~2.5\%, respectively. More generally, we establish a new analysis pipeline for the phase shift induced by neutrinos and other free-streaming dark radiation which additionally offers new avenues for exploring physics beyond the Standard Model in a signature-driven and model-agnostic way.

	\vskip10pt
	\hrule
	\vskip10pt
\end{titlepage}

\thispagestyle{empty}
\setcounter{page}{2}
\tableofcontents

\clearpage
\pagenumbering{arabic}
\setcounter{page}{1}
\section{Introduction}
\label{sec:introduction}

Cosmology and particle physics are deeply intertwined. Microscopic processes have been shaping our universe on small and large scales which allows us to use cosmological observations to test a wide range of theoretical scenarios. In fact, cosmic surveys provide a crucial avenue for probing particles with gravitational interactions, very weak couplings with the Standard Model~(SM) of particle physics or very small masses~(cf.~e.g.~\cite{Flauger:2022hie, Amin:2022soj, Green:2022hhj}). Moreover, precise measurements of the cosmic microwave background~(CMB), the large-scale structure~(LSS) of the universe and the abundances of light elements from big bang nucleosynthesis~(BBN) have now become precise enough to more generally complement terrestrial and astrophysical experiments~(see e.g.~\cite{Annis:2022xgg, Chang:2022lrw}). It is therefore interesting to deeply understand these data, the observables and the theoretical signatures to uncover exquisite information about nature and the fundamental laws of physics.\medskip

The neutrino is the second most abundant particle in the universe with a number density of about~\SI{112}{\per\centi\meter\cubed} per species for a total of approximately~\SI{336}{\per\centi\meter\cubed}~(only surpassed by the roughly 411~CMB~photons per cubic centimeter). These are cosmic neutrinos which were in thermal contact with the rest of the SM~particles in the very early universe. They decoupled from the primordial plasma about one second after the big bang at a temperature of approximately~\SI{1}{\mega\electronvolt} and have been freely streaming through the universe ever since, forming the cosmic neutrino background~(C$\nu$B; see e.g.~\cite{Scott:2024rwc} for a recent review). This background is however hard to directly detect in the laboratory~\cite{Bauer:2022lri}, e.g.\ as proposed by the Princeton Tritium Observatory for Light, Early-Universe, Massive-Neutrino Yield~(PTOLEMY)~\cite{PTOLEMY:2019hkd}, since neutrinos are only interacting via the weak force in the Standard Model and the~C$\nu$B has a thermal spectrum with a temperature of about~\SI{1.95}{\kelvin}. On the other hand, neutrinos are not only very abundant, but also make up about 41\%~of the radiation energy density and, therefore, effectively that fraction of the total energy density in the early universe. Even though gravity is the weakest force, the~C$\nu$B therefore leaves sizable gravitational imprints in our cosmological observables which we can use to infer the properties of these cosmic neutrinos.\medskip

In addition to photons and neutrinos, any particle beyond the Standard Model~(BSM) that is relativistic will also contribute to the radiation density. Such additional light particles are predicted in many BSM~scenarios, such as relativistic axion-like particles and other light scalars, light sterile neutrinos and dark photons~(see~\cite{Abazajian:2012ys, Marsh:2015xka, Hook:2018dlk, Green:2019glg, Allahverdi:2020bys, Asadi:2022njl, Dvorkin:2022jyg, Antel:2023hkf} for reviews). If these particles were in thermal contact with the primordial plasma at any point in the cosmic history, they will exhibit an appreciable number density and contribute to the energy density with an amount that would have already been or will be observable in the future. We can therefore gravitationally probe these light thermal relics in the same way as the cosmic neutrino background. In turn, we can use measurements of the radiation density of the universe to infer constraints on the interactions of these BSM~particles to the Standard Model~(see e.g.~\cite{Brust:2013ova, Chacko:2015noa, Baumann:2016wac, Ferreira:2018vjj, DEramo:2018vss, Arias-Aragon:2020shv, Ghosh:2020vti, Ferreira:2020bpb, Dror:2021nyr, Green:2021hjh, DEramo:2021usm, Caloni:2022uya, Bianchini:2023ubu, DEramo:2023nzt, Badziak:2024szg, Badziak:2024qjg, DEramo:2024jhn}). For simplicity, we will generally refer to neutrinos and the cosmic neutrino background in the following, but most statements in this paper also directly apply to light thermal BSM~particles.\medskip

Extensions of the Standard Model may also alter the interactions of neutrinos beyond the weak force. Neutrinos may, for instance, be coupled to each other directly, to dark matter or to new light BSM~particles, such as majorons~\cite{Berryman:2022hds}. Such interactions could modify the free-streaming behavior predicted for SM~neutrinos, which might in particular cause them to remain coupled to the primordial plasma longer than expected, to re-couple at later times or propagate in a fluid-like~state.\medskip

The gravitational interactions of neutrinos and other light relics affect our cosmological observables in distinct ways. We can in particular probe the following three properties of SM~neutrinos: (i)~their energy density~$\rho_\nu$, (ii)~their free-streaming nature, and (iii)~the sum of their masses~$\sum m_\nu$. In this paper, we focus on properties~(i) and~(ii), which are important in the early universe. The energy density of neutrinos is commonly parametrized by the effective number of relativistic~(free-streaming) degrees of freedom~$\Neff$ according to
\begin{equation}
	\Neff = \frac{8}{7} \left(\frac{11}{4}\right)^{\!4/3} \frac{\rho_\nu}{\rho_\gamma} \equiv a_\nu\hskip1pt\frac{\rho_\nu}{\rho_\gamma}\, ,	\label{eq:Neff}
\end{equation}
with the photon energy density~$\rho_\gamma$ and $a_\nu \approx 4.40$. In the Standard Model, the value of this parameter is $\Neff = \Neff^\mathrm{SM} = 3.044$ for the three neutrino species~\cite{Akita:2020szl, Froustey:2020mcq, Bennett:2020zkv, Cielo:2023bqp, Drewes:2024wbw, Drewes:2024nbg}, while BSM~physics may increase or decrease this value. A change in the radiation density alters the background evolution of the universe. In contrast, modifying the free-streaming nature, i.e.\ the free-streaming fraction or length, instead affects the perturbations of the primordial plasma.

\begin{figure}
	\centering
	\includegraphics{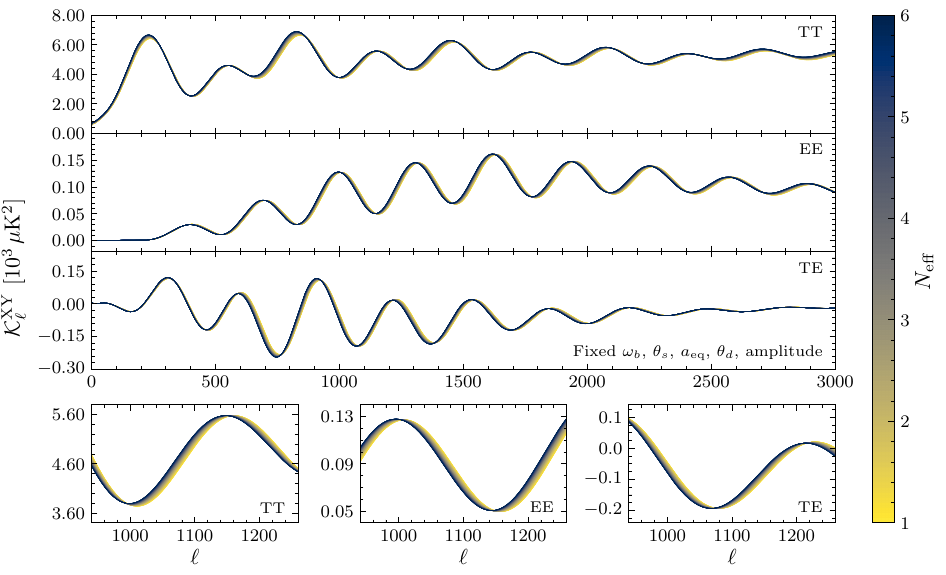}
	\caption{Illustration of the phase shift as imprinted in the temperature~(TT), polarization~(EE) and cross-correlation~(TE) power spectra with removed Silk damping,~$\mathcal{K}_\ell^{XY}$, as defined in~\eqref{eq:K_ell}. To highlight this effect as a function of the energy density of relativistic free-streaming species as parameterized by~$\Neff$, the following quantities were fixed as in~\cite{Follin:2015hya, Wallisch:2018rzj}: the physical baryon density~$\omega_b$, the scale factor at matter-radiation equality~$a_\mathrm{eq}$, the angular size of the sound horizon~$\theta_s$, the angular size of the damping scale~$\theta_d$ and the height of the fourth peak. The upper panels display the spectra over the observationally most relevant range of multipoles~$\ell$ for current CMB~experiments, while the lower panels zoom in around the fourth TT~peak to better exhibit the imprinted shift. This is the characteristic signature of free-streaming neutrinos and other light relics that we investigate and constrain in this paper.}
	\label{fig:Kl_phase_shift}
\end{figure}

Variations in the neutrino density or their free-streaming properties lead to specific signatures in our cosmological~BBN, CMB and LSS~measurements. We refer to the reviews~\cite{Lesgourgues:2013sjj, Lattanzi:2017ubx, Baumann:2018muz, Wallisch:2018rzj, Gerbino:2022nvz, Green:2022bre, Grohs:2023voo, Cooke:2024nqz} for further details. Here, we focus on the imprints in the cosmic microwave background. In the CMB~anisotropies, the main effects of relativistic species at the background level are on the sound horizon and the damping tail. Since the angular size of the sound horizon is very well measured, the main feature is a change in the Silk damping scale~\cite{Hou:2011ec}. At the perturbation level, free-streaming radiation induces an amplitude and a phase shift in the acoustic oscillations of the~CMB~\cite{Bashinsky:2003tk}. These manifest as a shift in the amplitude and location of the peaks and troughs in the temperature and polarization spectra for multipoles~$\ell \gtrsim 200$, as illustrated in Fig.~\ref{fig:Kl_phase_shift}.%
Schematically, this can be understood as follows: Free-streaming neutrinos move essentially at the speed of light while information in the photon-baryon fluid travels at the smaller speed of sound. This means that neutrinos effectively overtake the photons, which are tightly coupled to the baryons prior to recombination, and then gravitationally pull them slightly ahead of the sound horizon.\medskip

We focus on the phase shift in the acoustic oscillations since it is a distinct imprint of free-streaming radiation. While the amplitude of the CMB~peaks is modified as well, it also experiences other physical effects which means that disentangling those is more difficult and less direct. On the other hand, a phase is only imprinted in the acoustic oscillations either by free-streaming species or isocurvature perturbations~\cite{Baumann:2015rya}. Since we assume adiabatic initial conditions, a phase can only be induced by free-streaming radiation and is therefore an excellent signature to look for.\footnote{We note that various additional effects lead to shifts in the CMB~peak locations which can separately be accounted for and are distinct from a phase in the acoustic oscillations themselves~\cite{Pan:2016zla}.} In particular, the induced phase is constant for modes that entered the horizon deep in radiation domination while an increasingly smaller phase is imprinted on modes that entered the horizon at later times when the gravitational influence of radiation is suppressed~\cite{Bashinsky:2003tk, Baumann:2015rya}. This phase shift with its characteristic scale dependence was first detected directly in Planck~2013 temperature power spectrum in~\cite{Follin:2015hya}, and indirectly in Planck~2015 temperature and polarization spectra in~\cite{Baumann:2015rya}. Previous and more recent analyses of CMB~data related to free-streaming and interacting neutrinos~(and light BSM~particles) were reported in~\cite{Bell:2005dr, Friedland:2007vv, Cyr-Racine:2013jua, Wilkinson:2014ksa, Oldengott:2014qra, Brust:2017nmv, Lancaster:2017ksf, Choi:2018gho, Kreisch:2019yzn, Park:2019ibn, Forastieri:2019cuf, Ghosh:2019tab, Blinov:2020hmc, Das:2020xke, RoyChoudhury:2020dmd, Brinckmann:2020bcn, Ghosh:2021axu, Venzor:2022hql, Kreisch:2022zxp, Taule:2022jrz, RoyChoudhury:2022rva, Ge:2022qws, Brinckmann:2022ajr, Das:2023npl, Venzor:2023aka, Ange:2023ygk, Allali:2024cji, Ghosh:2024wva}. The same phase shift is naturally also imprinted in LSS~spectra~\cite{Baumann:2017lmt, Baumann:2017gkg, Green:2020fjb} and was constrained in the BOSS~DR12 dataset through a generalized analysis of the baryon acoustic oscillations~\cite{Baumann:2017gkg, Baumann:2019keh} which was recently also applied to data from the Dark Energy Spectroscopic Instrument~\cite{Whitford:2024ecj}. Additional related studies in the context of~LSS, BBN and gravitational waves include~\cite{Ghosh:2017jdy, Lague:2019yvs, Grohs:2020xxd, MoradinezhadDizgah:2021upg, Yeh:2022heq, Loverde:2022wih, Camarena:2023cku, He:2023oke, Lee:2023uxu, Camarena:2024daj, Bond:2024ivb}.

In this paper, we directly detect the phase shift from free-streaming neutrinos using two distinct approaches in current temperature and polarization datasets from the Planck satellite, the Atacama Cosmology Telescope~(ACT) and the South Pole Telescope~(SPT). In our first approach, we build on the work of~\cite{Follin:2015hya} and derive a template for the shifts in the peak locations of not only the temperature, but for the first time also the polarization power spectra~$C_\ell^{XY}$\hskip-2pt, $X,Y \in \{T, E\}$, which we refer to as the spectrum-based template~(SBT). In our second and new approach, we infer a template for the shift in the phase of the photon density perturbations~$\delta_\gamma$ which we refer to as the perturbation-based template~(PBT). To parameterize the signal in a physical way and subsequently estimate its size from data, we will introduce two new parameters in analogy to~$\Neff$ which we refer to as~$\Nmultipole$ for~SBT and~$\Nphase$ for~PBT. While the SBT~method effectively measures the phase as imprinted in the peaks and troughs of the CMB~spectra, the novel PBT~approach directly measures the phase induced by free-streaming neutrinos and other light relics for the first time. The implementation of the latter approach allows for a robust inference of the phase shift and additionally opens up new ways to search for physics beyond the Standard Model in a signature-driven and model-agnostic way.\bigskip

This paper is organized as follows: In Section~\ref{sec:background}, we provide a detailed explanation of the neutrino-induced phase shift and derive a theoretical template for its characteristic shape in the power spectra and the photon perturbations, respectively. We also validate and compare both methods in mock-data analyses. In Section~\ref{sec:analysis-forecasts}, we infer constraints on the size of the phase shift in terms of~$\Nmultipole$ and~$\Nphase$ from current CMB~data and forecast the projected sensitivity of future surveys. We conclude in Section~\ref{sec:conclusions}. A set of appendices contains additional details on the validation and implementation of the analysis pipeline~(Appendix~\ref{app:validation}), comprehensive summary of the constraints obtained from current CMB~data~(Appendix~\ref{app:tables}), and a full description of our forecasts and projected sensitivities for future surveys~(Appendix~\ref{app:forecasts}).

\section{Free-Streaming Neutrinos in the~CMB}
\label{sec:background}

Neutrinos are the lightest and most weakly coupled particles in the Standard Model which makes them hard to probe in the laboratory. Since these particles however were in thermal equilibrium with the rest of the primordial plasma at very early times, we can study their sizable gravitational influence in cosmology. In this work, we probe the SM~prediction that neutrinos have been freely streaming through the universe for a long time before the~CMB was released and, therefore, imprinted a phase shift in the acoustic oscillations that we see in the~CMB and~LSS. In this section, we first provide some background on the neutrino-induced phase shift~(\textsection\ref{sec:background-theory}), but also refer to~\cite{Bashinsky:2003tk, Follin:2015hya, Baumann:2015rya, Wallisch:2018rzj} for further details. We then describe how we numerically derive our templates for the shift effectively induced in the CMB~spectra~(SBT;~\textsection\ref{sec:spectrum-based-template}) and for the shift directly imprinted in the photon-baryon perturbations~(PBT;~\textsection\ref{sec:perturbation-based-template}), respectively. We finally compare and validate the templates on mock spectra in~\textsection\ref{sec:validation_comparison} to prepare for the data analysis in the next section.

\subsection{Theoretical Background}
\label{sec:background-theory}

All Standard Model particles were in thermal equilibrium at the high temperatures in the early universe. While the heavy particles become Boltzmann suppressed at some point, neutrinos are the first known species to decouple from the primordial plasma at a redshift $z_{\nu, \mathrm{dec}} \sim \num{e10}$, considerably earlier than hydrogen recombination and the subsequent decoupling of CMB~photons at $z_\mathrm{rec} \approx z_{\gamma, \mathrm{dec}} \approx 1080$. Since neutrinos effectively do not interact~(within the Standard Model) with themselves or any other particles after weak decoupling, their free-streaming length is very large. In other words, they have been freely streaming through the cosmos \mbox{at~(close to) the speed of light}.\medskip

The primordial perturbations imprinted at the beginning of the hot big bang induced sound waves in the primordial plasma propagating information at the speed of sound~$c_s \approx c/\sqrt{3}$. After neutrinos decouple, their perturbations travel at effectively the speed of light~$c$ which is significantly faster than the sound speed~$c_s$ of the photon-baryon fluid. As a result, their induced metric fluctuations extend beyond the sound horizon which then slightly distort the photon and baryon perturbations towards larger scales. This means that the acoustic oscillations in particular acquire a small phase~$\phi$, which schematically means for the photon/baryon density fluctuations~$\delta_{\gamma,b}$ with adiabatic initial conditions that
\begin{equation}
	\delta_{\gamma,b}(\vec{k}) \approx B_{\gamma,b}(\vec{k}) \cos(k r_s) \to \tilde{B}_{\gamma,b}(\vec{k}) \cos(k r_s + \phi)\, ,
\end{equation}
with wavenumber $k = |\vec{k}|$, size of the sound horizon~$r_s$ and $B_{\gamma,b} \to \tilde{B}_{\gamma,b}$ capturing the corresponding effect on the amplitude~$B_{\gamma,b}$ induced by the free-streaming neutrinos. For modes that entered the horizon deep in radiation domination~($k \to \infty$), the phase approaches a constant, $\phi \propto \epsilon(\Neff)$ at linear order in the fractional energy density~$\epsilon$ of neutrinos~\cite{Bashinsky:2003tk, Baumann:2015rya},
\begin{equation}
	\epsilon(\Neff) \equiv \frac{\rho_\nu}{\rho_r} = \frac{\Neff}{a_\nu + \Neff} \, ,	\label{eq:fractional_density}
\end{equation}
where $\rho_r = \rho_\gamma + \rho_\nu$ is the total radiation density and $a_\nu$~is defined in~\eqref{eq:Neff}. Since the relative contribution of neutrinos to the total energy density of the universe decreases with time as we move from deep inside the radiation-dominated era into the matter-dominated epoch, the phase~$\phi$ acquires a characteristic wavenumber dependence which smoothly goes to zero at least as fast as~$k^2$ as $k \to 0$~\cite{Baumann:2015rya}.\footnote{These asymptotic behaviors can also be anticipated in Figures~\ref{fig:template_SBT} and~\ref{fig:template_PBT} which show the phase shift as extracted at the level of the spectra and perturbations, respectively.}\medskip

When the CMB~photons were released, a snapshot of the sound waves was imprinted as the temperature anisotropies and polarization which we can observe today in our CMB~experiments. The two most prominent features directly visible in their power spectra are the acoustic oscillations and the exponential damping tail caused by photon diffusion prior to recombination. In the context of this paper, it is also worth mentioning that processes at $z < z_\mathrm{rec}$ alter the CMB~statistics before we measure them on Earth, in particular through the integrated Sachs-Wolfe~(ISW) effect and weak gravitational lensing.

A change in the neutrino~(and/or dark radiation) energy density affects the temperature and polarization spectra predominantly through a shift in the frequency of the acoustic oscillations and the damping scale of the tail~\cite{Hou:2011ec}. Since the first temperature peak is very well measured, we usually fix the angular size of the sound horizon~$\theta_s$ which implies that the spectra are more damped as we increase~$\Neff$. This effect is however degenerate with a variation in the primordial helium fraction~$Y_p$ since an increase in its value results in a smaller number density of free electrons which in turn leads to less damping. For fixed~$\theta_s$, the parameters~$\Neff$ and~$Y_p$ are therefore anti-correlated. Within the standard cosmological model~$\Lambda$CDM, we usually set the value of~$Y_p$ to be consistent with BBN~predictions as a function of the physical baryon density~$\omega_b$ and~$\Neff$.

The phase shift~$\phi$ in the acoustic oscillations induced by an increase~(decrease) in the free-streaming radiation manifests as a coherent displacement of the peak locations in both the temperature and polarization power spectra to larger~(smaller) scales. Since changes in other cosmological parameters can also affect the amplitude of the peaks, the shift in the amplitude associated with a change in~$\Neff$ is less distinct than the phase shift. This makes the phase shift a particularly robust and unique observable for free-streaming relativistic species.\medskip

\begin{figure}
	\centering
	\includegraphics{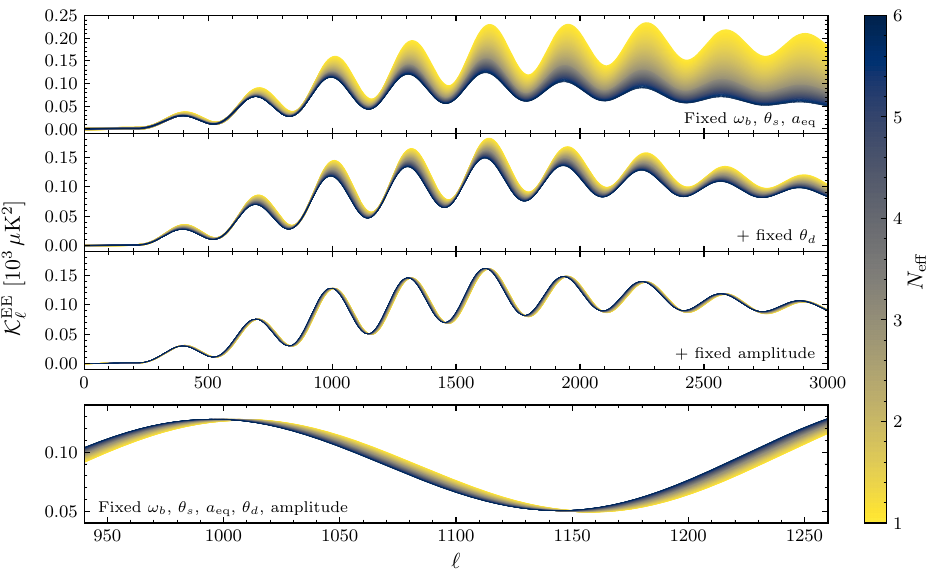}
	\caption{Variation of the lensed CMB~polarization power spectrum with removed Silk damping,~$\mathcal{K}^{EE}_\ell$, defined in~\eqref{eq:K_ell}, as a function of the energy density of relativistic free-streaming species as parameterized by~$\Neff$. To isolate the phase shift, we sequentially fix a number of parameters to their fiducial values. Following~\cite{Follin:2015hya, Wallisch:2018rzj}, the physical baryon density~$\omega_b$, the scale factor at matter-radiation equality~$a_\mathrm{eq}$ and the angular size of the sound horizon at recombination~$\theta_s$ are held fixed in all panels. In the top panel, the main effect is the variation of the damping scale~$\theta_d$. This is why we fix~$\theta_d$ in the second panel by adjusting the primordial helium fraction~$Y_p$ which reveals the amplitude shift as a free-streaming signature. In the third panel, we additionally normalize the amplitude of the spectra at the fourth peak such that the remaining variation is the multipole shift~$\delta\ell$. The bottom panel presents a smaller multipole range of the third panel to better highlight the shift towards larger angular scales.}
	\label{fig:KEE_phase_shift}
\end{figure}

\begin{figure}
	\centering
	\includegraphics{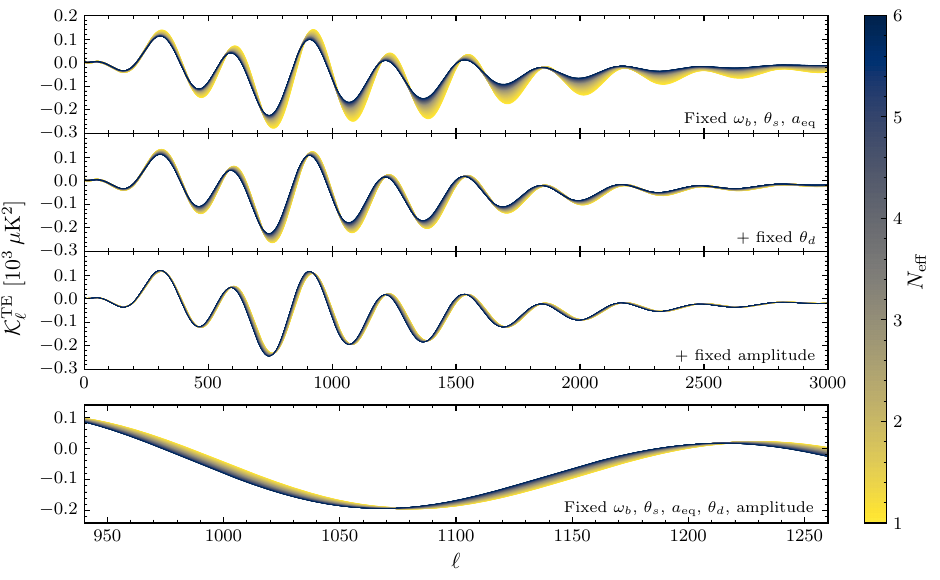}
	\caption{Same as Fig.~\ref{fig:KEE_phase_shift} for the lensed CMB~temperature-polarization cross-spectrum~$\mathcal{K}^{TE}_\ell$.}
	\label{fig:KTE_phase_shift}
\end{figure}

We illustrate the effects of free-streaming radiation on the CMB~spectra for E~modes and the TE~cross-correlation in Figures~\ref{fig:KEE_phase_shift}
and~\ref{fig:KTE_phase_shift}. By sequentially fixing a range of parameters, which are affected by a change in~$\Neff$, to their fiducial values, we isolate the phase shift as the remaining signature of free-streaming radiation.\footnote{In this way, Figures~\ref{fig:KEE_phase_shift} and~\ref{fig:KTE_phase_shift} also reproduce the upper and lower~EE and~TE panels of Fig.~\ref{fig:Kl_phase_shift}. We refer to~\cite{Follin:2015hya, Wallisch:2018rzj} for the equivalent figure for the temperature power spectrum.}
We display the undamped, lensed power spectrum~$\mathcal{K}_\ell^{XY}$ which is defined as
\begin{equation}
	\mathcal{K}_\ell^{XY} \equiv \frac{\ell(\ell+1)}{2\pi}\, C_\ell^{XY} \exp\left\{a(\theta_d\ell)^\kappa \right\},	\label{eq:K_ell}
\end{equation}
where~$\theta_d$ is the angular size of the damping scale, i.e.\ we approximately remove the Silk damping from the lensed CMB~spectrum~$C_\ell^{XY}$. Following~\cite{Wallisch:2018rzj}, we fit the damping parameters~$a$ and~$\kappa$ to the power spectra for the fiducial $\Lambda$CDM~cosmology\hskip1pt\footnote{We assume neutrinos to be massless throughout this paper. Given the current bounds on neutrino masses, this is a good approximation in the early universe which essentially does not affect the template derivation~\cite{Imig:priv} or the data analysis of the temperature and polarization spectra, which we also explicitly check in~\textsection\ref{sec:validation_comparison}.} defined in Table~\ref{tab:parameters}%
\begin{table}
	\centering
	\sisetup{group-digits=false}
	\begin{tabular}{l S[table-format=1.5] l}
			\toprule
		Parameter 				& {Fiducial Value}		& Description															\\
			\midrule[0.065em]
		$\omega_b$ 				& 0.02238 				& Physical density of baryons $\omega_b \equiv \Omega_b h^2$			\\
		$\omega_c$				& 0.12011				& Physical density of cold dark matter $\omega_c \equiv \Omega_c h^2$	\\
		$100\,\theta_s$ 		& 1.04178 				& $100\,\times\,$angular size of the sound horizon at decoupling 		\\
		$\ln(\num{e10}\As)$		& 3.0448				& Logarithm of the primordial scalar amplitude~(at~$k_\star$)			\\
		$\ns$					& 0.96605 				& Scalar spectral index~(at~$k_\star$)									\\
		$\tau$ 					& 0.0543 				& Optical depth due to reionization										\\
			\midrule[0.065em]
		$\Neff$					& 3.044 				& Effective number of free-streaming relativistic species				\\
		$\Nmultipole$			& 3.044 				& Effective number of multipole-shifting relativistic species\\
		$\Nphase$				& 3.044 				& Effective number of phase-shifting relativistic species	\\
		$Y_p$					& {`BBN'\hskip6.5pt}	& Primordial helium fraction 											\\
			\bottomrule
	\end{tabular}
	\caption{Parameters of the fiducial $\Lambda$CDM~model inspired by the Planck~2018 best-fit cosmology~\cite{Planck:2018vyg} and of the extensions considered in this work. The pivot scale is $k_\star = \SI{0.05}{\per\Mpc}$ and we assume massless neutrinos throughout this work, which is a good approximation in the early universe and in the context of our phase-shift measurements~(see~\textsection\ref{sec:validation_comparison} for a detailed discussion). The primordial helium fraction~$Y_p$ is generally fixed by consistency with~BBN, which implies $Y_p = 0.24534$ for this set of parameter values. We introduce the parameters~$\Neff^{\delta X}$, $X = \ell, \phi$, to set the amplitude of the spectrum-based and perturbation-based template, respectively.}
	\label{tab:parameters}
\end{table}
to effectively remove the exponential diffusion damping: $a \approx 0.68$ and $\kappa \approx 1.3$ for $\theta_d \approx \num{1.6e-3}$.\footnote{To be more precise, following~\cite{Wallisch:2018rzj}, we approximately remove the Silk damping according to~\eqref{eq:K_ell} by fitting the damping parameters~$a$ and~$\kappa$ to the peaks of the CMB~spectra using a least-squares method. We note that small variations in the values of the fitting parameters do not affect any results in this paper.}\medskip

We can isolate the characteristic phase shift in the temperature, polarization and cross-correlation power spectra following the procedure outlined in~\cite{Follin:2015hya, Wallisch:2018rzj} by subsequently fixing to their fiducial values those parameters that mainly govern or lead to the same effects as the change in~$\Neff$. We fix the physical baryon density~$\omega_b$, the scale factor at matter radiation equality $a_\mathrm{eq} \equiv \omega_m/\omega_r$ and the angular size of the sound horizon~$\theta_s$ to remove the effects on the frequency and radiation driving through a change in~$a_\mathrm{eq}$. We remove the remaining background effect on the angular size of the Silk damping scale~$\theta_d$ from changes in the expansion rate of the universe due to a different amount of radiation by adjusting the primordial helium fraction~$Y_p$ accordingly. This effectively isolates the impact of neutrino perturbations on the~CMB with the remaining amplitude and phase shifts. In the figures, we can in particular observe the multipole shift~$\delta\ell$ to smaller multipoles~(larger scales). In the following~(\textsection\ref{sec:spectrum-based-template}), we first numerically derive a template for~$\delta\ell$ based on the procedure laid out in this paragraph. In~\textsection\ref{sec:perturbation-based-template}, we adopt a more fundamental approach and directly isolate the phase shift at the level of the photon-baryon perturbations.

\subsection{Spectrum-Based Template}
\label{sec:spectrum-based-template}

We will now introduce one of the two distinct approaches of this work to detect the phase shift from free-streaming neutrinos in CMB~datasets. This is a spectrum-based method to characterize the phase shift induced by free-streaming neutrinos in the power spectra. This builds on the original framework developed in~\cite{Follin:2015hya} which uses a spectrum-based template to effectively shift the entire temperature and polarization spectra~$C_\ell^{XY}$, $X,Y \in \{T, E\}$, as imprinted by the phase shift of the acoustic oscillations in the CMB~peaks and troughs.

\subsubsection{Template Derivation from the Spectra}
\label{sec:spectrum-template}

\begin{figure}
	\centering
	\includegraphics{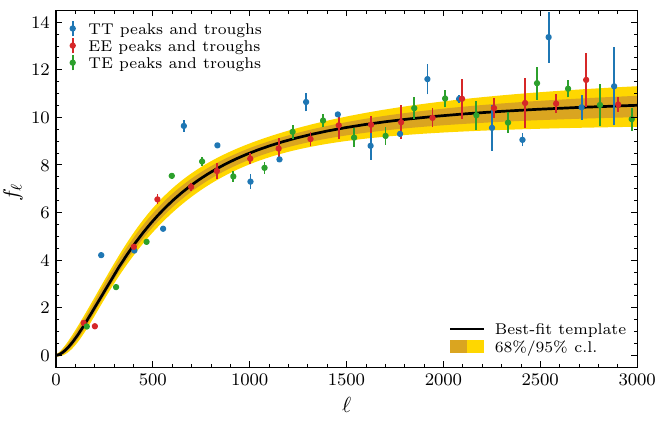}
	\caption{Template of the spectrum-based multipole shift~$f_\ell$ as defined in~\eqref{eq:delta_ell_def1}. The numerical phase shifts for the TT, EE and TE~spectra, which are displayed in \textcolor{tabblue}{blue}, \textcolor{tabred}{red} and \textcolor{tabgreen}{green}, respectively, were obtained from sampling 100~different cosmologies with varying number of relativistic species~$\Neff \in [1,6]$ and are normalized to the multipole shift from~3.044 to one species. The error bars indicate the standard deviation in these measurements at the respective peaks/troughs of the CMB~power spectra relative to the fiducial cosmology. The best fit for the spectrum-based template~$f_\ell$ of~\eqref{eq:f_ell_def} is shown in black and its corresponding~$1\sigma$~($2\sigma$)~confidence interval are shaded in dark~(light) \textcolor{goldenrod}{gold}.}
	\label{fig:template_SBT}
\end{figure}

The multipole shift induced by $\Neff$~free-streaming relativistic species in the acoustic peaks of the CMB~power spectra with respect to the Standard Model expectation with three free-streaming neutrinos can be parametrized as~\cite{Follin:2015hya}:
\begin{equation}
	\delta\ell(\Neff) = A(\Neff)\,f_\ell \, ,	\label{eq:delta_ell_def1}
\end{equation}
where~$A(\Neff)$ is the size of the effect and $f_\ell$~is the template function that encodes its multipole dependence. This is motivated by the physical insights discussed in~\textsection\ref{sec:background-theory} which in particular suggest that the size should linearly grow with the fractional neutrino density~$\epsilon(\Neff)$ of~\eqref{eq:fractional_density} at leading order in this quantity,
\begin{equation}
	A(\Neff) \equiv \frac{\epsilon(\Neff)-\epsilon(3.044)}{\epsilon(1)-\epsilon(3.044)}\, ,	\label{eq:ANeff}
\end{equation}
where we chose to normalize~$A(\Neff)$ to the shift between the Standard Model value of $\Neff^\mathrm{SM} = 3.044$ and one neutrino species.\footnote{The normalization of~$A(\Neff)$ is arbitrary and here set such that $\delta\ell(\Neff = 1) = f_\ell$. This is the same choice as in~\cite{Follin:2015hya} which we adopted for consistency and ease of comparison with their results. We also note that it is useful to compute the multipole shift relative to the fiducial cosmology with three free-streaming neutrinos and $\Neff = 3.044$ rather than the overall phase shift with respect to zero species and $\Neff = 0$ since we expect our measurement to be around the SM~value.} Those same insights about the expected behavior for modes that entered the horizon deep in the radiation- and matter-dominated eras, respectively, lead us to the following functional form of the spectrum-based template~$f_\ell$:
\begin{equation}
	f_\ell = \frac{\ell_\infty}{1 + (\ell/\ell_\star)^\xi}\, ,	\label{eq:f_ell_def}
\end{equation}
which asymptotes to the constant~$\ell_\infty$ for $\ell \to \infty$ and smoothly approaches zero governed by the parameters~$\ell_\star$ and $\xi < 0$ as $\ell \to 0$. (This functional form for the template was also employed in the large-scale structure analysis proposed in~\cite{Baumann:2017gkg} and performed in~\cite{Baumann:2019keh, Whitford:2024ecj}; see also~\cite{Green:2020fjb}). As illustrated in Fig.~\ref{fig:template_SBT},%
this fitting function well approximates the shifts in the CMB~peaks and troughs for $\ell_\infty = 11.0 \pm 0.6$, $\ell_\star = 483 \pm 53$ and $\xi = -1.69 \pm 0.13$.

To obtain this template, we followed the prescription in~\cite{Follin:2015hya} and numerically calculated the phase shift~$\delta\ell$ from the lensed spectra obtained using the Boltzmann solver \texttt{CLASS}~\cite{Blas:2011rf}.\footnote{We derived the template from lensed spectra to ensure self-consistency with the observed power spectra, which are lensed and will be employed in our data analysis. We also computed the fit using unlensed spectra and found only minor differences in the fitting parameters within the multipole range of interest, $\ell \lesssim 3000$.} Specifically, we compared the fiducial $\Lambda$CDM~model of Table~\ref{tab:parameters}\hskip1pt\footnote{We derive the template using this one fiducial cosmology, but have explicitly checked that it is stable against variations of the $\Lambda$CDM~parameter values within current observational constraints.} with 100~cosmologies of varying effective number of neutrino species $\Neff \in [1,6]$. To isolate the effect of free-streaming neutrinos in the CMB~spectra, we kept the physical baryon density~$\omega_b$, the scale factor at matter-radiation equality $a_\mathrm{eq}$, the sound horizon $\theta_s$ and the damping scale $\theta_d$ fixed across all models, equivalently to what we showed in the third panel of Fig.~\ref{fig:KEE_phase_shift}. We also removed contributions from the ISW~effect to prevent contamination from post-recombination anisotropies. Unlike~\cite{Follin:2015hya}, which used a logarithmic template, we adopted the physically motivated fitting formula~\eqref{eq:f_ell_def}. Additionally, to obtain the best fit for our template, we do not rely solely on the TT~spectrum as in~\cite{Follin:2015hya}, but also incorporate the polarization spectra~TE and~EE, which constrain more precisely the template shape, as shown in Fig.~\ref{fig:template_SBT}. This is because the polarization signal is a direct measure of the quadrupole moment at the surface of last scattering, with no contributions from effects such as the Sachs-Wolfe and integrated Sachs-Wolfe effects. As a result, the acoustic peaks in the polarization spectra provide a cleaner view of the primordial sound waves and appear sharper than those in the temperature spectrum.

\subsubsection[Multipole Shifting the Spectra with \texorpdfstring{$\Nmultipole$}{Neff-delta-ell}]{Multipole Shifting the Spectra with $\mathbf{N}_\mathbf{eff}^{\boldsymbol{\delta\ell}}$}
\label{sec:spectrum-method}

Having derived the spectrum-based template~\eqref{eq:f_ell_def}, we now proceed to incorporate it in the observables so that we can subsequently use it to constrain the multipole shift induced by free-streaming neutrinos with CMB~data. To facilitate this, we define a new parameter~$\Nmultipole$ which we refer to as the effective number of multipole-shifting relativistic species. While the standard parameter~$\Neff$ captures all effects of free-streaming relativistic particles on the~CMB at both the background and the perturbation level, the new parameter~$\Nmultipole$ only controls the size of the multipole shift. Since the power spectra computed using~\texttt{CLASS} intrinsically contain the shift inherent to~$\Neff$, we have to remove this shift before we introduce the shift through~$\Nmultipole$. This means that the multipole shift induced by $\Nmultipole$~such species relative to $\Neff$~species is given by
\begin{equation}
	\Delta\ell(\Nmultipole, \Neff) \equiv \delta\ell(\Nmultipole) - \delta\ell(\Neff) = \bigl[A(\Nmultipole) - A(\Neff)\bigr] f_\ell\, .	\label{eq:deltaellnu}
\end{equation}
In this way, we ensure that the correct size of the phase shift is consistently contained in our observables. This procedure in particular prevents an artificial multipole shift for the physical model of~$\Neff$~free-streaming species, which is defined by~$\Neff$ and~$\Nmultipole$ taking the same values, even when both parameters are independently varied in a data analysis.

Given our goal of robustly determining the number of free-streaming species from data, we want to emphasize that the value of~$\Neff$ may count not only free-streaming, but also other~(fluid-like) relativistic species in a standard CMB~analysis~(e.g.\ within~$\Lambda\mathrm{CDM} + \Neff$) since~$\Neff$ is predominantly sensitive to the damping tail.\footnote{Free-streaming and fluid-like radiation have the same effects at the level of the background energy density and only free-streaming species induce imprints at the perturbation level, which in particular means that fluid-like species do not induce a phase shift. Instead of directly measuring the phase shift, as we do, a new parameter such as~$N_\mathrm{fluid}$ can alternatively be introduced in the cosmological model to distinguish fluid-like from free-streaming species, cf.~e.g.~\cite{Bell:2005dr, Friedland:2007vv, Baumann:2015rya, Brust:2017nmv, Blinov:2020hmc, Brinckmann:2020bcn, Brinckmann:2022ajr}.} By contrast, our new parameter~$\Nmultipole$ definitively counts the number of free-streaming species since it only measures their characteristic multipole shift. We will explicitly illustrate this behavior when we validate our analysis pipeline in~\textsection\ref{sec:validation_comparison}.\medskip

To implement the multipole shift into the power spectra, we follow~\cite{Follin:2015hya} and incorporate it in the undamped component without the integrated-Sachs-Wolfe effect. To be precise, we first remove the exponential diffusion damping from the power spectra~$C_\ell$ to get the undamped spectra~$\mathcal{K}_\ell$ as defined in~\eqref{eq:K_ell}. For the TT~and TE~spectra, we then decompose these undamped spectra into their ISW~component, \mbox{non-ISW}~component and the cross-term between these two, $\mathcal{K}_\ell = \mathcal{K}^\mathrm{ISW}_\ell + \mathcal{K}^{\cancel{\mathrm{ISW}}}_\ell + \mathcal{K}^\mathrm{cross}$. To include the multipole shift~$\Delta\ell$, we only shift the multipoles $\ell \to \ell + \Delta\ell$ in the ISW-free component,
\begin{equation}
	\mathcal{K}^{\cancel{\mathrm{ISW}}}_\ell \rightarrow \mathcal{K}^{\cancel{\mathrm{ISW}}}_{\ell + \Delta\ell}\, .	\label{eq:deltaell_data}
\end{equation}
Finally, we multiply these terms by the original damping term to arrive at the shifted CMB~power spectra~$\tilde{C}_\ell = C^\mathrm{ISW}_\ell + C^{\cancel{\mathrm{ISW}}}_{\ell + \Delta\ell} + C_\ell^\mathrm{cross}$. This prescription will allow us to set up an analysis pipeline in Section~\ref{sec:analysis-forecasts} and to test the robustness of the physical model independently of~$\Neff$ by directly fitting for the multipole shift via~$\Nmultipole$.

\subsection{Perturbation-Based Template}
\label{sec:perturbation-based-template}

We have introduced the spectrum-based template to parameterize the phase shift induced by free-streaming neutrinos and defined the new parameter~$\Nmultipole$, which controls the size of the phase shift independently of other effects of neutrinos or other light species according to the template function~$f_\ell$ of~\eqref{eq:f_ell_def}. While this approach offers a reliable way to constrain the phase shift with CMB~data, it remains approximate since it only captures the effective imprint of free-streaming neutrinos in the power spectra. At a fundamental level, however, a shift in the photon-baryon perturbations is induced directly by the neutrino perturbations as they propagate to the surface of last scattering and then to us. In this sense, the effective spectrum-based description overlooks certain complexities such as the projection effects arising from the fact that multiple $k$~modes in the photon perturbations contribute to a single $\ell$~mode in the CMB~spectra. Similarly, due to the finite width of the last scattering surface, the power spectra receive contributions from a narrow but finite time interval, rather than a single time slice at recombination~\cite{Hu:1997hp, Pan:2016zla}. Additionally, in the context of~SBT, an important limiting factor is the gravitational lensing of CMB~photons, caused by the matter distribution between the last scattering surface and us. This lensing process smooths the acoustic oscillations by redistributing power among neighboring $\ell$~modes, making it increasingly difficult to identify the CMB~peaks and, in turn, extract the phase shift template, especially at high multipoles $\ell \gtrsim 3000$. The combination of these effects may distort the direct relationship between the phase shift at the perturbation level and its manifestation in the angular CMB~power spectra observed in our surveys~(cf.~\cite{Pan:2016zla}).\medskip

Given these limitations, it is worthwhile to explore a more fundamental method that directly modifies the time-dependent perturbations in the photon-baryon fluid. Acknowledging that we do not expect significant differences between the two approaches with current data, the perturbation-based method not only provides a more accurate representation of the underlying physics, but will also enhance our ability to probe the free-streaming nature of neutrinos with future high-precision~CMB~data.

\subsubsection{Template Derivation from the Perturbations}
\label{sec:perturbation-template}

\begin{figure}
	\centering
	\includegraphics{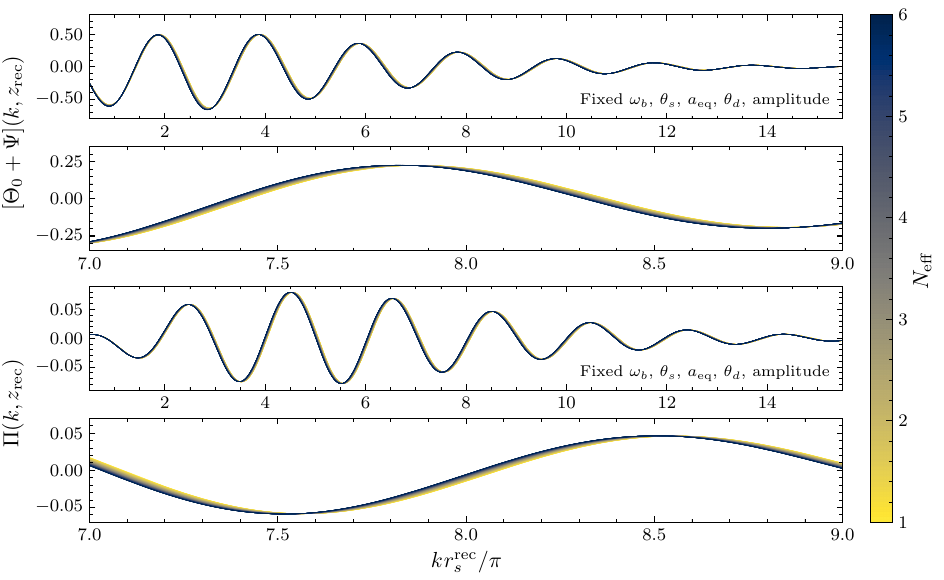}
	\caption{Variation of the perturbations in the photon-baryon fluid at recombination as a function of the amount of free-streaming radiation as parameterized by~$\Neff$. We show the effect on the monopole Sachs-Wolfe term~$\Theta_0 + \Psi$ in the top two panels and the polarization field~$\Pi$ in the bottom two panels for $\Neff \in [1, 6]$ as indicated by the colorbar. The second and fourth panels display a much smaller range of wavenumbers for better visualization of the phase shift in these quantities. As for the multipole shift illustrated in Figures~\mbox{\ref{fig:Kl_phase_shift}--\ref{fig:KTE_phase_shift}}, the perturbations are normalized at the fourth peak with~$\omega_b$, $a_\mathrm{eq}$, $\theta_s$ and~$\theta_d$ held fixed, such that the remaining \mbox{variation is the phase shift~$\delta\phi$.}}
	\label{fig:perturbation_phase-shift}
\end{figure}

Neutrino perturbations induce a phase shift in the fluctuations in the photon-baryon plasma. Of relevance for the CMB~power spectra, they primarily affect the dependence of two key quantities on mode~$k$ and evolution in conformal time~$\eta$: the Sachs-Wolfe term,~$\Theta_0 + \Psi$, and the strength of the polarization field~$\Pi$. Here, $\Theta_0$~represents the monopole temperature perturbations associated with the photon density fluctuations,~$\Theta_0 = \delta_\gamma/4$, while $\Psi$~is the Newtonian potential and accounts for the temperature fluctuations induced by the gravitational redshift of photons. The polarization field~$\Pi$ quantifies the quadrupole moment of the photon distribution, which directly determines the degree of CMB~polarization. Importantly, the induced phase shift is the same for both terms since the polarization field is related to the SW~term by a total time derivative, $\Pi \sim \frac{d}{d\eta}\!\left(\Theta_0+\Psi\right)$, which does not alter the phase-shift structure~\cite{Pan:2016zla}. We show this explicitly in Fig.~\ref{fig:perturbation_phase-shift}%
as a function of~$\Neff$. Here, following the same prescription of~\textsection\ref{sec:spectrum-based-template}, we fixed~$\omega_b$, $a_\mathrm{eq}$, $\theta_s$ and~$\theta_d$ across all models to isolate the effect of free-streaming neutrinos.\medskip

\begin{figure}
	\centering
	\includegraphics{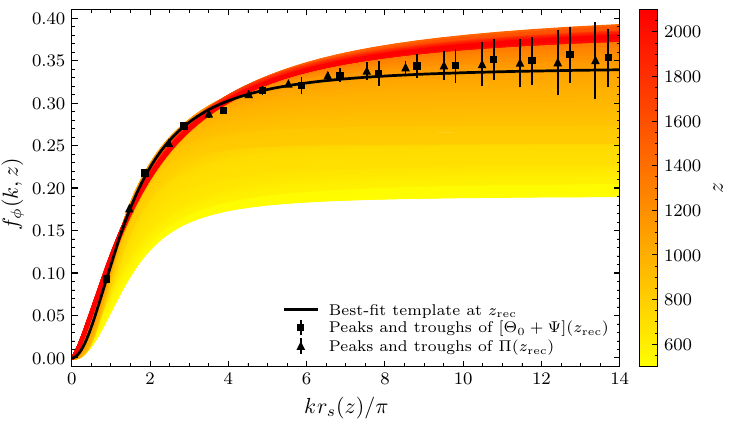}\vspace{-5pt}
	\caption{Perturbation-based template~$f_\phi$, as defined in~\eqref{eq:f_phi_def}, as a function of scale~$k r_s$ and redshift~$z$. In black squares and triangles, we display the numerical phase shifts at recombination for~$\Theta_0 + \Psi$ and~$\Pi$, respectively, as shown in Fig.~\ref{fig:perturbation_phase-shift}. These were obtained from sampling 100~cosmologies with varying number of free-streaming species~$\Neff \in [1, 6]$, with the normalization chosen to correspond to the phase shift from~3.044 to one species. The black line shows the best fit to the squares and triangles for the template~\eqref{eq:f_phi_def} at recombination,~$f_\phi(k, z_\mathrm{rec})$. The best-fitting templates at all other redshifts $z \in [500, 2100]$ are color coded as indicated by the colorbar.}
	\label{fig:template_PBT}
\end{figure}

Given all of these considerations, we parameterize the phase shift induced by free-streaming neutrinos in the perturbations of the photon-baryon fluid in a similar way as for the multipole shift~$\delta\ell$ of~\eqref{eq:delta_ell_def1}, but with two important differences. First, the characteristic shape of the phase shift is a function of wavenumber~$k$ instead of multipole~$\ell$. Second, the size of the phase shift changes with redshift~$z$ as the universe transitions from being radiation dominated to being matter dominated which suppresses the gravitational effect of free-streaming neutrinos. We have to accurately take this redshift evolution into account in contrast to~\eqref{eq:delta_ell_def1}, which is independent of redshift, because the CMB~spectra are snapshots at the time of recombination. We therefore describe the phase shift induced by $\Neff$~free-streaming relativistic species with respect to the Standard Model expectation with three active neutrinos as
\begin{equation}
	\delta\phi(k, z; \Neff) = A(\Neff)\,f_\phi(k,z)\, ,	\label{eq:delta_phi_def1}
\end{equation}
with the normalization~$A(\Neff)$ provided in~\eqref{eq:ANeff}. The perturbation-based template~$f_\phi(k,z)$ is well approximated by the fitting function
\begin{equation}
	f_\phi(k,z) = \frac{\phi_\infty}{1 + [k r_s /(k r_s)_\star]^\xi}\, ,	\label{eq:f_phi_def}
\end{equation}
where~$r_s = r_s(z)$ is the size of the comoving sound horizon at redshift~$z$, and the parameters~$\phi_\infty$, $(k r_s)_\star$ and~$\xi$, which define the shape and scale dependence of the template, are themselves functions of~$z$. In Figure~\ref{fig:template_PBT},%
we present this template across the range of redshifts $z \in [500, 2100]$ which is most relevant to the CMB~power spectra. To aid visualization of how the best-fit parameters were obtained, the black squares and triangles represent the phase shift for the SW~term and the polarization field, respectively, at the time of recombination. The solid black line shows the corresponding fitted template at this time. For further insight, we show the evolution of the best-fit parameters and their~$1\sigma$ uncertainties as a function of redshift in Fig.~\ref{fig:PBT_scatter_vs_z} in Appendix~\ref{app:validation}. Near the most relevant redshifts close to recombination, we find that the parameters take values of approximately $\phi_\infty \approx 0.11\pi \approx 0.34$, $(k r_s)_\star \approx 1.5\pi$ and $\xi \approx -2.0$.\footnote{We can compare the fitted value for~$\phi_\infty$ with the analytic expectation. To facilitate this comparison, we have to adjust the normalization of~\eqref{eq:ANeff} to be between three and zero species resulting in $\tilde{\phi}_\infty \approx 0.15\pi\hskip0.5pt\epsilon(\Neff)$. This is about 20\% smaller than the theoretically calculated value at linear order in~$\epsilon(\Neff)$ of~$\tilde{\phi}_\infty^\mathrm{th} \approx 0.191\pi\hskip0.5pt\epsilon(\Neff)$ from~\cite{Bashinsky:2003tk, Baumann:2015rya} which is consistent with the expected corrections from higher orders in~$\epsilon(\Neff)$.}

To obtain these templates, we followed the same prescription as outlined above and numerically calculated the phase shift~$\delta\phi$ directly from the SW~term and polarization field as computed by~\texttt{CLASS}. The resulting templates, which we show in Fig.~\ref{fig:template_PBT}, align with our expectations. Specifically, the phase shift always asymptotes to a constant at large values of~$k r_s$, while it approaches zero as $k r_s \to 0$. Furthermore, the overall size of the phase shift in the photon perturbations decreases with redshift which is a consequence of photons freely streaming after their decoupling.

\subsubsection[Phase Shifting the Perturbations with \texorpdfstring{$\Nphase$}{Neff-delta-phi}]{Phase Shifting the Perturbations with $\mathbf{N}_\mathbf{eff}^{\boldsymbol{\delta\phi}}$}
\label{sec:sources_to_Cells}

With this semi-analytic prescription of the neutrino-induced phase shift in hand, it is important to clarify how it translates into a measurable shift in the multipoles of the CMB~power spectra. To understand this connection, we first recall the relationship between the physical source terms, which arise from perturbations in the photon-baryon fluid, and the angular power spectra:
\begin{equation}
	C^{XY}_\ell = \frac{2}{\pi} \int\! \d k\, k^2 \Delta^2_{\mathcal{R}}(k) \,\Delta^X_\ell\hskip-1pt(k)\Delta^Y_\ell\hskip-1pt(k)\, ,	\label{eq:Cell_from_transfer}
\end{equation}
where~$X,\,Y$ refer to the temperature~$T$ and polarization mode~$E$, and~$\Delta^2_{\mathcal{R}}(k)$ is the primordial curvature power spectrum. The transfer functions~$\Delta^X_\ell(k)$ can be written as a line-of-sight integral of the physical source term~$S_X(k,\eta)$ and the geometric projection factor~$P_{X\ell}(k[\eta_0-\eta])$,
\begin{equation}
	\Delta^X_\ell(k) = \int_0^{\eta_0}\! \d\eta\,S_X(k,\eta)\,P_{X\ell}(k[\eta_0-\eta])\, .	\label{eq:transfers_from_sources}
\end{equation}
where $\eta_0$ is the conformal time today and $P_{X\ell}(k[\eta_0-\eta])$~represents the projection from Fourier to harmonic space, typically expressed as a combination of spherical Bessel functions. The source term~$S_X(k,\eta)$ encodes the physical processes in the photon-baryon fluid, including the Sachs-Wolfe term and polarization field, both of which are directly affected by neutrino perturbations through their characteristic phase shift. In Newtonian gauge, the polarization source term takes the form~\cite{Hu:1997hp}
\begin{align}
	S_E(k,\eta) = \frac{\sqrt{6}}{2} g_\gamma(\eta)\, \Pi(k,\eta)\, ,	\label{eq:SE}
\end{align}
with the photon visibility function~$g_\gamma(\eta) \equiv \tau^\prime \ee^{-\tau}$, which sharply peaks at the surface of last scattering and is defined in terms of the optical depth~$\tau = \tau(\eta)$ and its derivative~$\tau^\prime = \d\tau/\d\eta$. Evidently, the time and mode-dependent phase shift in the polarization field~$\Pi(k,\eta)$ caused by neutrino perturbations, shown in Fig.~\ref{fig:perturbation_phase-shift}, leads to a corresponding shift in the multipoles of the EE~power spectrum, as described by~\mbox{\eqref{eq:Cell_from_transfer}--\eqref{eq:SE}}.

The same principle, though more intricate, applies to the temperature source term, which receives contributions from multiple physical processes that distinctly affect the evolution of the perturbations. As a result, it is common to express the integral for the transfer function as a sum of the individual sources,~$S_{T,0}$, $S_{T,1}$ and~$S_{T,2}$, weighted by the appropriate projection operator given by the spherical Bessel functions~$j_\ell$ and their first and second derivatives, respectively~\cite{Hu:1997hp}:
\begin{equation}
	\begin{split}
		\Delta^T_\ell(k) = \int_0^{\eta_0}\! \d\eta\,\bigl\{	& S_{T,0}(k,\eta)\, j_{\ell}(k[\eta_0-\eta]) + S_{T,1}(k,\eta)\, j^\prime_{\ell}(k[\eta_0-\eta])							\\
																&\!+ S_{T,2}(k,\eta) \left[3 j^{\prime\prime}_{\ell}(k[\eta_0-\eta]) + j_{\ell}(k[\eta_0-\eta])\right]\!/2 \bigr\}\, ,
	\end{split} \label{eq:transferT}
\end{equation}
with the respective source terms
\begin{align}
	S_{T,0}(k,\eta) &= \underbrace{g_\gamma\left(\Theta_0+ \Psi\right)}_{\text{SW}} \underbrace{+g_\gamma\left(\Phi-\Psi\right) +2 \ee^{-\tau} \Phi^\prime}_{\text{ISW}} \underbrace{+1/k^2\left(g_\gamma\,\theta_b^\prime+g_\gamma^\prime\,\theta_b\right)}_{\text{Doppler}}\, ,	\\[-1pt]
	S_{T,1}(k,\eta) &= \underbrace{\ee^{-\tau}k \left(\Psi-\Phi\right)}_{\text{ISW}}\, , 						\\
	S_{T,2}(k,\eta) &= \underbrace{\frac{\sqrt{6}}{2}g_\gamma\, \Pi}_{\text{polarization}}\, ,
\end{align}
where we suppressed the arguments on the right-hand side for simplicity, $\Phi$~is the spatial curvature perturbation and $\theta_b$~is the baryon velocity divergence. We also highlighted the different contributions of the~SW, ISW, Doppler and polarization terms.

As discussed in~\textsection\ref{sec:perturbation-template}, free-streaming species induce the phase shift in the SW~term and the polarization field~(see Fig.~\ref{fig:perturbation_phase-shift}), while the ISW~contribution is not affected. In addition, the baryon velocity divergence~$\theta_b$ of course also experiences the same phase shift prior to the decoupling of the baryons from the photons. After this decoupling, the redshift evolution of the phase shifts in the baryon and photon perturbations slightly differ. We are however able to ignore this small difference and model the phase shift in~$\theta_b$ in the same way as in the SW~and polarization terms since the Doppler contribution to the temperature transfer function~$\Delta^T_\ell(k)$ is subdominant and the difference in the time evolution only occurs at times when the visibility function further suppresses the signal~(see Appendix~\ref{app:validation} for additional details).\medskip

Building on these foundations, we aim to investigate the size of the neutrino-induced phase shift in CMB~data by introducing a shift directly in the $k$~modes of the relevant physical source terms of the photon-baryon perturbations. To achieve this, we define the effective number of phase-shifting relativistic species~$\Nphase$. This new parameter independently controls the phase shift based on the perturbation-based template~$f_\phi$ in analogy to~$\Nmultipole$ in the spectrum-based method described in~\textsection\ref{sec:spectrum-method}. The $k$-mode shift~$\Delta\phi/r_s$ induced by $\Nphase$~such species relative to $\Neff$~species is then given by
\begin{equation}
	\Delta\phi(k, \eta; \Nphase, \Neff) \equiv \delta\phi(k, \eta; \Nphase) - \delta\phi(k, \eta; \Neff) = \left[A(\Nphase) - A(\Neff)\right] f_\phi(k, \eta)\, . \label{eq:delta_phi_def2}
\end{equation}
To shift the perturbations, we introduce~$\Delta\phi$ in the relevant terms of the temperature and polarization sources as
\begin{equation}
	\begin{split}
		\left[\Theta_0 + \Psi\right]\!(k, \eta) 	&\rightarrow \left[\Theta_0 + \Psi\right]\!(k + \Delta\phi/r_s, \eta)\, ,	\\
		\Pi(k, \eta) 								&\rightarrow \Pi(k + \Delta\phi/r_s, \eta)\, ,								\\
		\left[\theta_b/k\right]\!(k, \eta) 			&\rightarrow \left[\theta_b/k\right]\!(k + \Delta\phi/r_s, \eta)\, ,		\\
		\left[\theta^\prime_b/k^2\right]\!(k, \eta) &\rightarrow \left[\theta^\prime_b/k^2\right]\!(k + \Delta\phi/r_s,\eta)\, .	
	\end{split}	\label{eq:deltaphi_data}
\end{equation}
Through the source and transfer functions, the phase shift~\eqref{eq:delta_phi_def2} ultimately translates into the observable multipole shift in the corresponding power spectra~\eqref{eq:Cell_from_transfer} as explained above. Importantly, the prescription in~\eqref{eq:deltaphi_data} ensures, similar to the SBT~approach, that no artificial shift is introduced for the physical model of~$\Neff$~free-streaming neutrinos, for which $\Neff = \Nphase$. This method therefore again enables a robust test for the presence of a phase shift in the CMB~spectra, independent of~$\Neff$, through the introduction of the new parameter~$\Nphase$.

\subsection{Mock-Data Validation of the Templates}
\label{sec:validation_comparison}

Before confronting observed CMB~data, it is important to validate our newly introduced spectrum- and perturbation-based methodologies for directly measuring the neutrino-induced phase shift and to compare them to verify their consistency. To this end, we use our new parameters~$\Nmultipole$ and~$\Nphase$ for the SPT~and PBT~method, respectively, and infer their values from mock data. We conducted Markov Chain Monte Carlo~(MCMC) analyses on temperature and polarization power spectra generated for a Planck-like experiment with the same specifications as adopted in~\cite{Allison:2015qca}. These analyses employed the publicly available sampler~\texttt{MontePython}~\cite{Audren:2012wb, Brinckmann:2018cvx} and a modified version of~\texttt{CLASS} that incorporates the prescriptions laid out in~\eqref{eq:deltaell_data} and~\eqref{eq:deltaphi_data} for the spectrum- and perturbation-based method, respectively.\medskip

We considered three mock datasets, each generated based on the fiducial $\Lambda$CDM~cosmology of Table~\ref{tab:parameters} with additional relativistic degrees of freedom exhibiting distinct physical properties. The purpose of these analyses is to ensure the robustness and reliability of the techniques, which we developed in this work to measure the phase shift, across various physical scenarios on CMB~data. Specifically, we created three mock datasets with the following characteristics: (1)~three massless, free-streaming SM~neutrinos corresponding to $\Neff = 3.044$, (2)~three free-streaming neutrinos with $\Neff = 3.044$ and a nonzero total mass $\sum m_\nu = \SI{0.06}{\electronvolt}$,\footnote{To be exact, we use two massless neutrinos and one massive neutrino with $m_\nu = \SI{0.06}{\electronvolt}$, and settings chosen to ensure that $\Neff = 3.044$ and $\sum m_\nu /\omega_\nu = \SI{93.14}{\electronvolt}$, where $\omega_\nu \equiv \Omega_\nu h^2$ is the physical density of neutrinos. While at least two of the three neutrinos are known to have different, nonzero masses, this simplified approach is sufficient for our purposes of testing the assumption of massless neutrinos in the context of our phase-shift measurements.} and (3)~three massless relativistic species, consisting of two free-streaming particles~($\Neff = 2$) and one fluid-like specie~\mbox{($N_\mathrm{fluid} = 1$)}.\footnote{To incorporate fluid-like relativistic species in our mock data, we used the interacting dark matter-dark radiation module of~\texttt{CLASS}, which is based on the ETHOS~framework~\cite{Cyr-Racine:2015ihg}, imposing fluid-like behavior on the dark radiation and setting the interacting dark matter abundance to a negligible level.}

Our analysis of case~(1) validates that both of our methods can correctly infer the size of the neutrino-induced phase shift. This in particular tests for potential biases when estimating the values of the phase-shift parameters~$\Nmultipole$ and~$\Nphase$, and verifies the consistency between the spectrum- and perturbation-based methodologies. Case~(2) investigates whether the presence of a mass introduces any systematic bias in the recovered phase shift and, therefore, tests our assumption of massless neutrinos in the rest of the paper. Finally, case~(3) evaluates the ability of~$\Nmultipole$ and~$\Nphase$ to effectively constrain free-streaming relativistic species in a mixed scenario, while also examining any discrepancies between the two approaches under these more complex conditions. In all cases, we vary the six baseline parameters of the standard $\Lambda$CDM~model, $\{\omega_b,\,\omega_c,\,\theta_s,\,\ln(\num{e10}\As),\,\ns,\,\tau\}$, as well as~$\Neff$ and either~$\Nmultipole$ or~$\Nphase$, depending on whether the spectrum- or perturbation-based method is applied. Flat priors are assumed for all sampled cosmological parameters, and the Markov chains are run until convergence is achieved, ensuring $R < 0.01$ according to the Gelman-Rubin criterion~\cite{Gelman:1992zz}.\medskip

\begin{figure}
	\centering
	\includegraphics{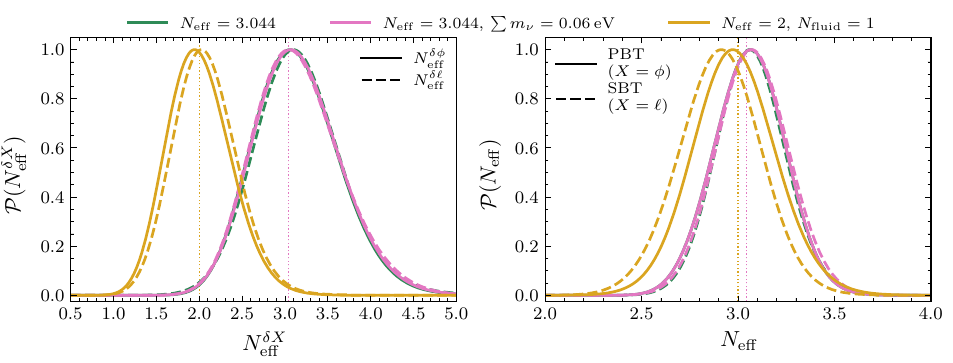}
	\caption{One-dimensional, normalized posterior distributions for~$\Neff^{\delta X}$, $X = \{\ell, \phi\}$~(\textit{left}), and~$\Neff$~(\textit{right}) for the three mock datasets described in the text. The case of three massless and free-streaming neutrinos with $\Neff = 3.044$ is shown in~\textcolor{tabgreen}{green}, while the \textcolor{tabpink}{pink} curves represent the case with three free-streaming neutrinos with $\Neff = 3.044$ and a nonzero total mass $\sum m_\nu = \SI{0.06}{\electronvolt}$. These two scenarios with a zero and nonzero neutrino mass are virtually indistinguishable, as expected, with the green curves barely visible underneath the pink posteriors. The third case with two free-streaming~($\Neff = 2$) and one fluid-like species~($N_\mathrm{fluid} = 1$) is included in~\textcolor{goldenrod}{gold}. Solid and dashed lines correspond to the perturbation-~($\Nphase$) and spectrum-based~($\Nmultipole$) method, respectively. Thin vertical dotted lines indicate the fiducial values used to generate the mock datasets in their respective colors. Overall, the figure highlights the robustness of our pipeline in recovering the phase shift and the consistency between the spectrum- and \mbox{perturbation-based approach}.}
	\label{fig:mock_planck}
\end{figure}

We display the resulting one-dimensional posterior distributions\hskip1pt\footnote{Throughout this work, we present all one-dimensional posterior distributions normalized to unity.} for~$\Neff^{\delta X}$, $X = \{\phi, \ell\}$, and~$\Neff$ for the three mock-data variations in Fig.~\ref{fig:mock_planck}.\footnote{For the interested reader, we also display the two-dimensional joint constraints on~$\Neff$ and~$\Nmultipole$ or~$\Nphase$ for the three mock datasets under consideration in Fig.~\ref{fig:mock_planck_2d} in Appendix~\ref{app:validation}.}%
Overall, this demonstrates that our pipeline consistently recovers the induced phase shift in the CMB~spectra with high precision across all mock datasets, demonstrating its robustness. This is underscored by the close alignment between the fiducial values of the input mock cosmologies and the obtained posterior distributions. In addition, the spectrum- and perturbation-based methods yield consistent results, confirming the reliability of both approaches.

The results on the second mock dataset, which contains three neutrinos with a nonzero total mass $\sum m_\nu = \SI{0.06}{\electronvolt}$, are virtually indistinguishable from the case of three massless, free-streaming SM~neutrinos in case~(1). This demonstrates that the inclusion of neutrino masses does not introduce a systematic bias in the constraints on the phase shift. Such an outcome is expected since neutrinos with such small masses are still relativistic at recombination. In addition, it validates our approach for the template derivation, which assumed massless neutrinos.

Finally, the mock dataset of case~(3) with two free-streaming and one fluid-like relativistic species, which is shown in gold in the figure, intuitively illustrates the key advantage of incorporating~$\Nmultipole$ or~$\Nphase$ into an analysis pipeline. We see that the inferred values are $\Neff^{\delta X} \approx 2$, which is the expected value, and $\Neff \approx 3$, which corresponds to the total number of relativistic species in the dataset even though the parameter is defined to only count the number of relativistic free-streaming particles. This is because the constraining power on~$\Neff$ is dominated by its effect on the background evolution as seen in the CMB~damping tail which does not discriminate between free-streaming and fluid-like particles. Consequently, its posterior distribution remains centered roughly around three, reflecting the total number of relativistic species present. In contrast, $\Nmultipole$~and~$\Nphase$ exclusively measure the induced phase shift, which allows us to correctly identify that only two of the three species are free-streaming. This simple example underscores the strength of our analysis pipeline in isolating the unique imprint of free-streaming relativistic species. By disentangling the phase shift effect from the total radiation energy density, our framework provides a robust probe for identifying the nature of relativistic species which may offer valuable insights into BSM~physics, including potential non-standard neutrino interactions or couplings to a dark sector.

\section{Data Analyses and Forecasts}
\label{sec:analysis-forecasts}

Having established the reliability of our methodology through our validation on mock datasets, we now proceed to investigate the size of the neutrino-induced phase shift in observed CMB~data. To this end, we employ the same analysis methodology as in the validation stage, i.e.\ we perform MCMC~analyses using the publicly available sampler~\texttt{MontePython} and a modified version of the Boltzmann code~\texttt{CLASS} that implements the prescriptions in~\eqref{eq:deltaell_data} and~\eqref{eq:deltaphi_data} for the spectrum- and perturbation-based method, respectively.\footnote{We note that our inferred constraints on the phase shift are insensitive to the particular choice of the fiducial cosmological parameters used to derive the templates given the correspondingly small error bars of the $\Lambda$CDM~parameters.} We infer the value of either~$\Nmultipole$ or~$\Nphase$ and consider different cases by either fixing or varying the effective number of relativistic free-streaming species~$\Neff$ and/or the primordial helium fraction~$Y_p$. This allows us to both delineate the various effects of these parameters on the observables and investigate the robustness of the phase shift: as discussed, both~$\Neff$ and~$\Neff^{\delta X}$ induce a phase shift while~$Y_p$ is degenerate with the background effect of~$\Neff$ on the damping tail. When held fixed, we set~$\Neff$ to its SM~value, i.e.~$\Neff^\mathrm{SM} = 3.044$, while~$Y_p$ is taken to be consistent with the predictions of~BBN. In all cases, we always marginalize over the six baseline parameters of the standard $\Lambda$CDM~model and assume flat~(linear) priors for all sampled cosmological parameters. We ensure convergence of the chains by requiring the Gelman-Rubin criterion $R < 0.01$.\medskip

We consider the following datasets in our analyses:
\begin{itemize}
	\item Planck~2018~(PR3): We use the Planck~2018~PR3 high-$\ell$~($\ell\geq 30$) \texttt{plik-lite} likelihood~\cite{Planck:2019nip} which contains TT~measurements up to $\ell_\mathrm{max}=2508$ and TE~and EE~measurements up to $\ell_\mathrm{max}=1996$.\footnote{We note that \texttt{plik-lite} provides a compressed and faster version of the \texttt{plik}~likelihood, with the parameters describing foregrounds and residual systematics already marginalized over~\cite{Planck:2019nip}. We checked explicitly that this choice does not affect our results by comparing \texttt{plik-lite} and \texttt{plik} constraints for the case where~$\Neff$, $Y_p$ and either~$\Nmultipole$ or~$\Nphase$ are varied simultaneously together with the $\Lambda$CDM~parameters finding excellent agreement.} For low multipoles in the range $2 \leq \ell \leq 29$, we employ the \texttt{commander} and \texttt{SimAll}~likelihoods for the TT~and EE~spectra, respectively. When using all these data products together, we label it as `P18~TTTEEE'. We also define `P18~TT' to identify the combination that includes only TT~information in the high-$\ell$ likelihood, while still incorporating all the low-$\ell$ information from the TT~and EE~spectra.
	
	\item ACT \& SPT: We use the \texttt{actpollite\_dr4}~\cite{ACT:2020frw} and \texttt{spt3g\_y1} likelihoods~\cite{SPT-3G:2022hvq} from the ACT and SPT~collaborations, respectively. The ACT~DR4 data release provides TT~measurements for $600 < \ell < 4126$, and TE~and EE~measurements in the range $350 < \ell < 4126$. The SPT-3G~2018 data instead include TT~measurements for $750 < \ell \leq 3000$, with TE~and EE~spectra covering $300 < \ell \leq 3000$. We use these data only in conjunction with the full Planck likelihood~(P18~TTTEEE),\footnote{We had independently checked that the datasets are consistent within the extended cosmologies analyzed in this work.} labeling this combination as `P18 + ACT + SPT'. To avoid correlations with the Planck~TT power spectrum, we follow~\cite{ACT:2020frw} and remove any overlap with ACT~DR4~TT measurements up to $\ell = 1800$.
	
	\item Planck~2021~(PR4): We use the latest Planck~2021~PR4 likelihood code, based on the NPIPE~data release~\cite{Planck:2020olo}, and consider both independent implementations:
	\begin{enumerate}
		\item \texttt{CamSpec}: We replace the Planck~PR3 \texttt{plik} \mbox{high-$\ell$} TTTEEE~likelihood with the more recent PR4~\texttt{CamSpec} likelihood~\cite{Efstathiou:2019mdh, Rosenberg:2022sdy}, while retaining the same \mbox{low-$\ell$}~likelihoods from the \texttt{plik}~combination, namely \texttt{commander} and \texttt{SimAll} for the TT~and EE~spectra, respectively. We label this combination `P21~[\texttt{CamSpec}]'.
		\item \texttt{HiLLiPoP} + \texttt{LoLLiPoP}: In this implementation, we replace both the \texttt{plik} \mbox{high-$\ell$} TTTEEE~likelihood with the PR4~\texttt{HiLLiPoP} likelihood~\cite{Tristram:2020wbi,Tristram:2023haj} and the \texttt{SimAll} \mbox{low-$\ell$}~EE~likelihood with the PR4~\texttt{LoLLiPoP} likelihood~\cite{Tristram:2020wbi,Tristram:2023haj}. The \texttt{commander} likelihood remains in use for the low-$\ell$ TT~spectra. We refer to this combination as `P21~[\texttt{HiLLiPoP}]'.
	\end{enumerate}
	We include Planck~2021~(PR4) to assess in~\textsection\ref{sec:comparison-data} how the recent updates in the NPIPE~data release compared to the official Planck~2018~(PR3) likelihood impact the constraints for both of our methods to measure the phase shift since small-scale noise and systematics were significantly reduced in the new frequency maps~\cite{Planck:2020olo, Rosenberg:2022sdy, Tristram:2023haj}.
\end{itemize}

\subsection{Spectrum-Based Analysis}
\label{sec:spectrum-analysis}

\begin{figure}
	\centering
	\includegraphics{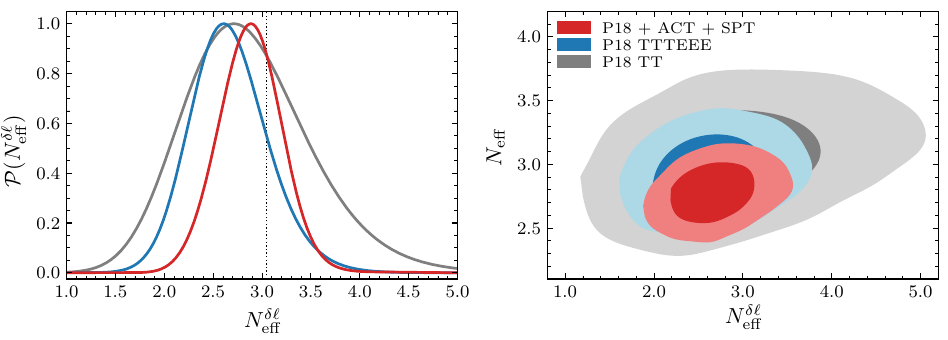}
	\caption{Constraints on the size of the phase shift inferred from current CMB~data in our spectrum-based analysis. \textit{Left:}~One-dimensional, normalized posterior distributions of~$\Nmultipole$~(with fixed $\Neff = \Neff^\mathrm{SM} = 3.044$) for the Planck TT-only results~(\textcolor{tabgray}{gray}), Planck~TT, TE and EE~likelihoods~(\textcolor{tabblue}{blue}), and when including~ACT and SPT~data~(\textcolor{tabred}{red}). \textit{Right:}~Two-dimensional posteriors of~$\Neff$ and~$\Nmultipole$ at~68\% and 95\%~confidence levels~(c.l.)\ for the same set of data as the left panel. Regardless of whether we vary or fix~$\Neff$, all distributions show an exclusion of $\Nmultipole = 0$ at high significance, in particular when including polarization data.}
	\label{fig:template_SBT_Ndell-Neff_analysis}
\end{figure}

We unequivocally detect the existence of the neutrino-induced phase shift in our spectrum-based analysis of observed CMB~data, as shown in Fig.~\ref{fig:template_SBT_Ndell-Neff_analysis}.%
This employs the multipole shift~\eqref{eq:deltaellnu} with the spectrum-based template~\eqref{eq:f_ell_def} as parameterized by~$\Nmultipole$ on three combinations of Planck~2018, ACT and~SPT data. In the left panel of Fig.~\ref{fig:template_SBT_Ndell-Neff_analysis}, we present the one-dimensional posterior distributions of~$\Nmultipole$ for the $\Lambda\mathrm{CDM} + \Nmultipole$~model, with~$\Neff$ fixed at~3.044 and $Y_p$~fixed to be consistent with the predictions of~BBN. In its right panel, we display the two-dimensional joint constraints on~$\Nmultipole$ and~$\Neff$, when we allow both of these parameters to vary.\medskip

In all cases, we find that CMB~data are consistent with the standard physical model of three free-streaming neutrinos, $\Neff = \Nmultipole = 3.044$, at the $1\sigma$~level. Compared to the first phase-shift detection with Planck~2013 TT-only data in~\cite{Follin:2015hya}, we find a sizable reduction in the uncertainties even for~`P18~TT' and especially with added polarization data, as expected. We also see that the non-Gaussianity of the posterior distributions, which originates from the nonlinear relationship between~$\Neff$ and the size of the phase shift~$A(\Neff)$, cf.~\eqref{eq:ANeff}, decisively shrinks with increased precision. More generally, $\Nmultipole = 0$ is excluded at~$7\sigma$, $10\sigma$ and~$14\sigma$ significance for~`P18~TT', `P18~TTTEEE' and `P18 + ACT + SPT', respectively,\footnote{In light of the non-Gaussian shape of the posterior distributions, we quantified the exclusion significance using likelihood ratios between the analyses of the $\Lambda\mathrm{CDM} + \Nmultipole$~model and a $\Lambda\mathrm{CDM}$~model with fixed $\Nmultipole = 0$. We find that the minimum~$\chi^2$ is lower by~24, 52 and~97 in the $\Lambda\mathrm{CDM} + \Nmultipole$~model for the three dataset combinations, which can be translated into approximate Gaussian significances of~$7\sigma$, $10\sigma$ and~$14\sigma$ for `P18~TT', `P18~TTTEEE' and `P18 + ACT + SPT', respectively.\label{fn:significance}} strongly indicating that the $\Neff$~relativistic degrees of freedom associated with neutrinos present in the early universe are predominantly free-streaming. Even the scenario of two free-streaming and one fluid-like species is excluded at a level of more than~$2.5\sigma$ for the `P18 + ACT + SPT' data combination. This can also be seen in the first two rows of Table~\ref{tab:N_deltaell_analysis},%
\begin{table}
	\centering
	\begin{tabular}{l c c c}
			\toprule
															& P18 TT										& P18 TTTEEE									& P18 + ACT + SPT								\\
			\midrule[0.065em]
		$\Lambda\mathrm{CDM} + \Nmultipole$ 				& $2.87^{+0.54\, (+1.37)}_{-0.73\, (-1.26)}$	& $2.67^{+0.39\, (+0.83)}_{-0.43\, (-0.81)}$	& $2.87^{+0.31\, (+0.66)}_{-0.35\, (-0.65)}$	\\[4pt]
		$\Lambda\mathrm{CDM} + \Nmultipole + \Neff$ 		& $2.82^{+0.58\, (+1.49)}_{-0.82\, (-1.35)}$	& $2.64^{+0.38\, (+0.85)}_{-0.45\, (-0.84)}$	& $2.68^{+0.29\, (+0.69)}_{-0.35\, (-0.64)}$	\\[4pt]
		$\Lambda\mathrm{CDM} + \Nmultipole + \Neff + Y_p$	& $2.76^{+0.59\, (+1.61)}_{-0.86\, (-1.44)}$	& $2.62^{+0.36\, (+0.86)}_{-0.46\, (-0.81)}$	& $2.68^{+0.31\, (+0.69)}_{-0.35\, (-0.67)}$	\\[2pt]
			\bottomrule
	\end{tabular}
	\caption{Overview of the spectrum-based constraints at 68\%~(95\%)~c.l.\ on the neutrino-induced phase shift as parameterized by~$\Nmultipole$ for the three dataset combinations described in Sec.~\ref{sec:analysis-forecasts}. We show the constraints for the three extended cosmologies with varying~$\Nmultipole$ considered in this work: we fix $\Neff = \Neff^\mathrm{SM}$ and $Y_p$~to be consistent with~BBN in the first row while we marginalize over~$\Neff$, and over both~$\Neff$ and~$Y_p$ in the subsequent rows. In all cases, we also marginalize over the six baseline parameters of the standard $\Lambda$CDM~model. We refer to Table~\ref{tab:sbt_all} in Appendix~\ref{app:tables} for a complete list of the mean values and their corresponding $1\sigma$~uncertainties.}
	\label{tab:N_deltaell_analysis}
\end{table}
which summarize the constraints on~$\Nmultipole$ for the aforementioned dataset combinations within the one- and two-parameter extensions of the~$\Lambda$CDM~model in which we vary~$\Nmultipole$ while either fixing or marginalizing over~$\Neff$.

In agreement with previous studies~\cite{Baumann:2015rya, Brust:2017nmv, Kreisch:2019yzn, Forastieri:2019cuf, Blinov:2020hmc, Das:2020xke, Brinckmann:2020bcn}, we also find that the inclusion of polarization data leads to a significant improvement in the constraints on~$\Nmultipole$, due to the sharper E-mode peaks which enhance the detectability of the phase shift. When we incorporate data from ground-based experiments, namely~ACT and~SPT, the obtained constraints sharpen by an additional 25\%~in precision, primarily due to the measurement of higher multipoles.

When we consider the `P18 + ACT + SPT' dataset, we observe two minor effects compared to the analyses with only the P18~data. First, when~$\Neff$ is fixed, the mean value for~$\Nmultipole$ shifts upward from~$2.7$ to~$2.9$. In contrast, when~$\Neff$ is allowed to vary, this upward movement in~$\Nmultipole$ is lessened, with the mean value decreasing by approximately~7\%, back to~$2.7$. Notably, this is the only combination of datasets for which marginalizing over~$\Neff$~(instead of fixing it to the SM~value) has any effect on the constraint on~$\Nmultipole$. The reason for these movements is likely due to the combination of the ACT~and SPT~datasets which show a $1.6\sigma$~difference in their respective constraints on~$\Neff$~\cite{ACT:2020gnv, SPT-3G:2022hvq}. This discrepancy in turn can impact the constraints on~$\Nmultipole$ differently depending on whether~$\Neff$ is varied or fixed, since marginalizing over~$\Neff$ also infers the total energy density in relativistic species and the amplitude shift associated with free-streaming particles from the data.

\begin{figure}
	\centering
	\includegraphics{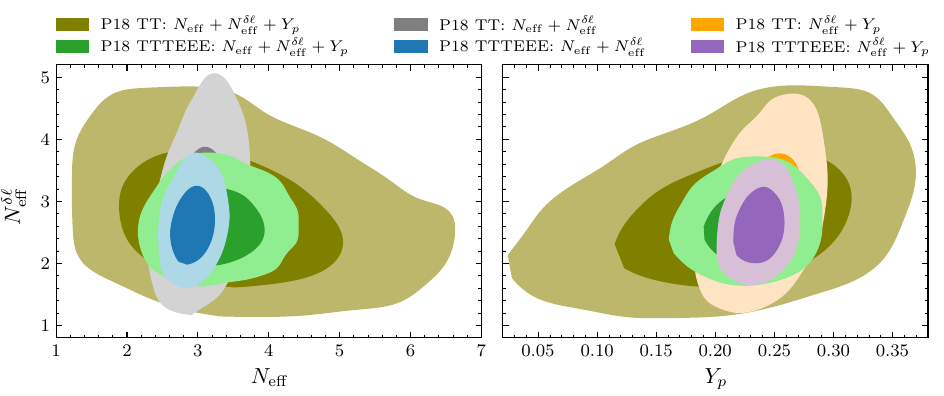}
	\caption{Effects of the primordial helium fraction~$Y_p$ on the spectrum-based analysis for the phase shift. The two panels show the two-dimensional posteriors on~$\Neff$ and~$\Nmultipole$~(\textit{left}), and the corresponding results for~$Y_p$ and~$\Nmultipole$~(\textit{right}), respectively. The displayed contours represent the 68\% and 95\%~confidence levels, derived from Planck TT-only, and TT, TE and EE~likelihoods, with the different colors indicating the analyzed datasets and extended cosmologies. As expected, we lose significant constraining power on~$\Neff$ when varying~$Y_p$, especially if we only consider temperature data. Instead, the~$\Nmultipole$ constraints do not depend on whether we use a cosmology with varying or fixed~$Y_p$, and significantly improve when polarization data are included.}
	\label{fig:template_SBT_Ndell-Neff-YHe_analysis}
\end{figure}

Finally, we turn to a comparison of the Planck constraints on~$\Neff$ and~$\Nmultipole$ for fixed or varying primordial helium fraction~$Y_p$ in Fig.~\ref{fig:template_SBT_Ndell-Neff-YHe_analysis},%
with the constraints on~$\Nmultipole$ being summarized in the second and third row of Table~\ref{tab:N_deltaell_analysis}, respectively. As expected, varying~$Y_p$, particularly when considering temperature data alone, significantly weakens the constraints on~$\Neff$, which we can clearly observe when comparing the olive and gray contours along~$\Neff$, with the former being inferred with varying~$Y_p$ and the latter with fixed~$Y_p$. As we have discussed, this is due to the degenerate effect of these parameters on the TT~damping tail. In contrast, for both `P18~TT' and `P18~TTTEEE', the constraint on~$\Nmultipole$ remains unaffected by varying either~$\Neff$ or~$Y_p$ which can be deduced by comparing the same olive and gray contours along~$\Nmultipole$, for example. This highlights the minimal degeneracy between~$\Nmultipole$ and~$\Neff$, and~$\Nmultipole$ and~$Y_p$. The lack of correlation can be explained by the fact that the phase shift induces a distinct feature in the power spectra that is difficult to mimic by other cosmological parameters. This finding is consistent with previous studies, which have highlighted the phase shift as a unique and reliable imprint to constrain~$\Neff$ since it helps to break degeneracies with other cosmological parameters~\cite{Baumann:2015rya, Brust:2017nmv, Kreisch:2019yzn, Forastieri:2019cuf, Blinov:2020hmc, Das:2020xke, Brinckmann:2020bcn, Ge:2022qws}. In this context, it is worth noting that $\Nmultipole$~exhibits its only significant degeneracy with the angular size of the sound horizon~$\theta_s$. Their strong correlation arises because these parameters effectively measure the phase and frequency of the acoustic oscillations over a limited number of peaks and troughs. Overall, these spectrum-based results emphasize the important role of the phase shift as a robust probe of the free-streaming nature of relativistic degrees of freedom.

\subsection{Perturbation-Based Analysis}
\label{sec:perturbation-analysis}

\begin{figure}
	\centering
	\includegraphics{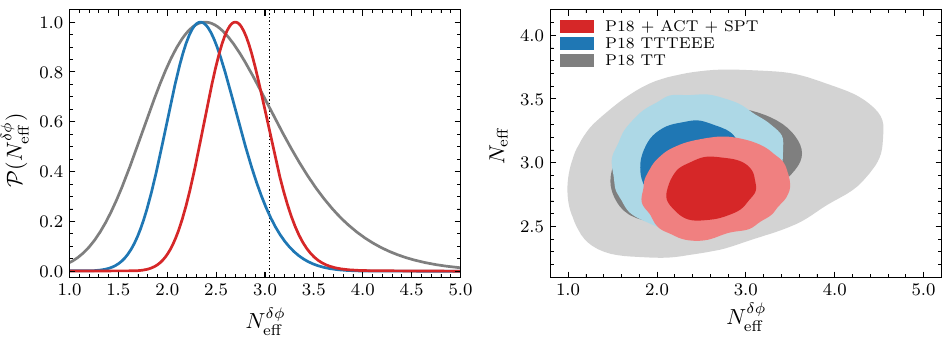}
	\caption{Constraints on the size of the phase shift inferred from current CMB~data in our perturbation-based analysis. \textit{Left:}~One-dimensional, normalized posterior distributions of~$\Nphase$~(with fixed $\Neff = \Neff^\mathrm{SM} = 3.044$) for the Planck TT-only results~(\textcolor{tabgray}{gray}), the Planck~TT, TE and EE~likelihood~(\textcolor{tabblue}{blue}), and when including ACT~and SPT~data~(\textcolor{tabred}{red}). \textit{Right:}~Two-dimensional posteriors of~$\Neff$ and~$\Nphase$ at~68\% and 95\%~c.l.\ for the same set of data as the left panel. For all likelihoods, we again exclude $\Nphase = 0$ at high significance and our results are consistent with the $\Nmultipole$~analyses, cf.~Fig.~\ref{fig:template_SBT_Ndell-Neff_analysis}.}
	\label{fig:PBT_Ndphi-Neff_analysis}
\end{figure}

\begin{figure}
	\centering
	\includegraphics{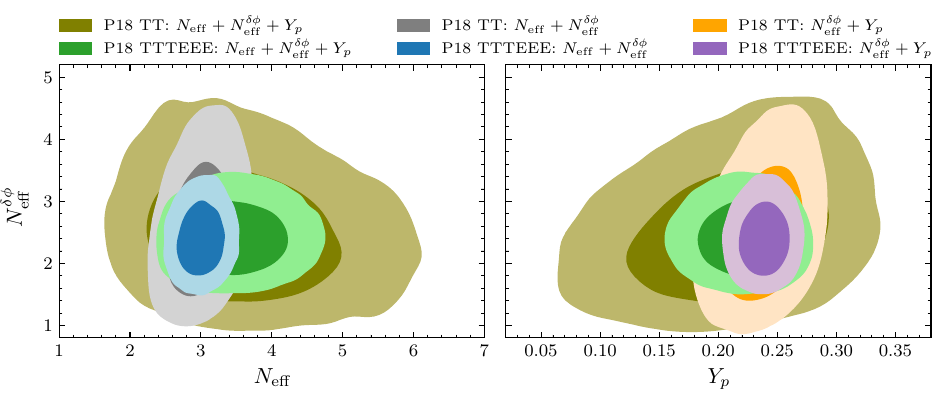}
	\caption{Effects of the primordial helium fraction~$Y_p$ on the perturbation-based analysis for the phase shift. This figure is analogous to Fig.~\ref{fig:template_SBT_Ndell-Neff-YHe_analysis} for the spectrum-based analysis and contains the same datasets, cosmological models and format. Similar to the results of this other method, we observe the importance of including polarization data and the well-known degeneracy~(lack thereof) between~$\Neff$~($\Nphase$) and~$Y_p$, respectively.}
	\label{fig:PBT_Ndphi-Neff-YHe_analysis}
\end{figure}

The results of the analyses, which employ the phase shift~\eqref{eq:delta_phi_def2} with the perturbation-based template~\eqref{eq:f_phi_def}, are displayed in Figures~\ref{fig:PBT_Ndphi-Neff_analysis}%
and~\ref{fig:PBT_Ndphi-Neff-YHe_analysis}.%
The former presents the one-dimensional posterior distributions of~$\Nphase$ for the $\Lambda\mathrm{CDM} + \Nphase$~model and the two-dimensional joint constraints on~$\Nphase$ and~$\Neff$, while the latter displays a comparison of the Planck constraints on~$\Neff$ and~$\Nphase$ for both fixed and varying~$Y_p$. A summary of the constraints on~$\Nphase$ derived from all of these analyses is also provided in Table~\ref{tab:N_deltaphi_analysis}.%
\begin{table}
	\centering
	\begin{tabular}{l c c c}
			\toprule
														& P18 TT										& P18 TTTEEE									& P18 + ACT + SPT								\\
			\midrule[0.065em]
		$\Lambda\mathrm{CDM} + \Nphase$					& $2.63^{+0.53\, (+1.49)}_{-0.83\, (-1.36)}$	& $2.41^{+0.34\, (+0.79)}_{-0.42\, (-0.73)}$	& $2.71^{+0.30\, (+0.65)}_{-0.34\, (-0.64)}$	\\[4pt]
		$\Lambda\mathrm{CDM} + \Nphase + \Neff$ 		& $2.59^{+0.58\, (+1.49)}_{-0.81\, (-1.34)}$	& $2.41^{+0.33\, (+0.78)}_{-0.42\, (-0.74)}$	& $2.62^{+0.30\, (+0.66)}_{-0.35\, (-0.64)}$	\\[4pt]
		$\Lambda\mathrm{CDM} + \Nphase + \Neff + Y_p$	& $2.50^{+0.57\, (+1.51)}_{-0.83\, (-1.36)}$	& $2.42^{+0.35\, (+0.78)}_{-0.42\, (-0.76)}$	& $2.62^{+0.30\, (+0.66)}_{-0.35\, (-0.63)}$	\\[2pt]
			\bottomrule
	\end{tabular}
	\caption{Overview of the perturbation-based constraints at 68\%~(95\%)~c.l.\ on the neutrino-induced phase shift as parameterized by~$\Nphase$ for the three dataset combinations described in Sec.~\ref{sec:analysis-forecasts}. This table follows the same format as Table~\ref{tab:N_deltaell_analysis}, with rows corresponding to the different cosmologies analyzed with fixed or varying~$\Neff$ and~$Y_p$. For a complete list of the obtained mean values and their $1\sigma$~uncertainties, we refer to Table~\ref{tab:pbt_all} in Appendix~\ref{app:tables}.}
	\label{tab:N_deltaphi_analysis}
\end{table}
These figures and table for the PBT~analysis are analogous to Fig.~\ref{fig:template_SBT_Ndell-Neff_analysis}, Fig.~\ref{fig:template_SBT_Ndell-Neff-YHe_analysis} and Table~\ref{tab:N_deltaell_analysis} for the SBT~analysis.\medskip

Overall, the main conclusions from this analysis are practically the same as those reached for the spectrum-based analysis described in~\textsection\ref{sec:spectrum-analysis}. We in particular find strong evidence for a non-zero phase shift in the CMB~anisotropies, i.e.\ $\Nphase \neq 0$, detected at~$6\sigma$, $10\sigma$ and~$14\sigma$ for~`P18~TT', `P18~TTTEEE' and `P18 + ACT + SPT', respectively.\footnote{We follow the same strategy as in our SBT~approach~(see footnote~\ref{fn:significance}) and determine the exclusion significance using the relevant likelihood ratios. Here, we find that the minimum~$\chi^2$ decreases by~18, 48 and~90 for~`P18~TT', `P18~TTTEEE' and `P18 + ACT + SPT', respectively.} In addition, the data are generally consistent with the standard physical model, $\Neff = \Nphase = 3.044$, at roughly the $1\sigma$~level~(see Fig.~\ref{fig:PBT_Ndphi-Neff_analysis}). Moreover, we again observe that the inclusion of polarization data substantially improves the constraints, while the addition of higher-multipole data from the ground-based experiments further tightens them, though the impact is again more modest. Finally, this analysis again confirms the robustness of the phase shift signature as a probe of free-streaming relativistic species since we find negligible correlation between the constraints on~$\Nphase$ and \mbox{those on~$\Neff$ and~$Y_p$~(see Fig.~\ref{fig:PBT_Ndphi-Neff-YHe_analysis}).}

As expected, these conclusions mirror those from the spectrum-based method, as the two approaches are in fact consistent with one another, as we explicitly showed on mock data in~\textsection\ref{sec:validation_comparison}. There are, however, some minor differences in the central values of the inferred posterior distributions for the two methods, particularly when considering Planck data alone, and we discuss them in more detail in the following.

\subsection{Comparison of the Analyses}
\label{sec:comparison-data}

As anticipated above, the conclusions drawn from the spectrum- and perturbation-based analyses are largely consistent, with only minor discrepancies between the central values of~$\Nmultipole$ and~$\Nphase$. While these differences are always well below the $1\sigma$~level, we find that the spectrum-based method using Planck~2018~(PR3) data systematically yields slightly larger values for the phase shift compared to the perturbation-based method. We now further investigate this difference by employing the Planck~2021~(PR4)~data.\medskip

Given our rigorous validation of both approaches using mock data in~\textsection\ref{sec:validation_comparison}, we can confidently exclude the possibility that these differences arise from a bias in either methodology. Instead, they must reflect a subtle feature in the datasets used for our analyses. In fact, we find that these discrepancies are most pronounced when using Planck data alone, as clearly shown in Tables~\ref{tab:N_deltaell_analysis} and~\ref{tab:N_deltaphi_analysis}, which suggests that they stem from features in this dataset. We confirm this intuition by analyzing the combination of ACT~and SPT~data independently and finding that the obtained constraints on~$\Nmultipole$ and~$\Nphase$ are in excellent agreement.

To further investigate this minor inconsistency, we turn to the latest Planck~2021~(NPIPE) data release~\cite{Planck:2020olo}, which features significant improvements in the processing of the time-ordered data, resulting in substantially lower noise levels on small angular scales compared to earlier Planck releases. Specifically, we leverage the two updated likelihood configurations~\cite{Efstathiou:2019mdh, Rosenberg:2022sdy, Tristram:2020wbi, Tristram:2023haj}, which we refer to as `P21~[\texttt{Camspec}]' and `P21~[\texttt{HiLLiPoP}]', to determine whether the observed discrepancies persist.\medskip

\begin{figure}
	\centering
	\includegraphics{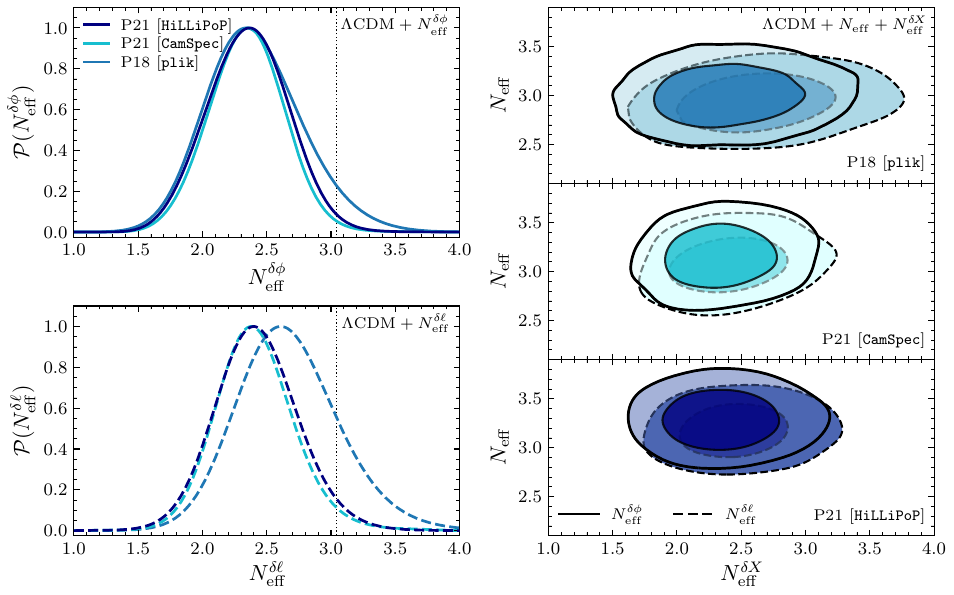}\vspace{-5pt}
	\caption{Impact of the updated Planck~2021 analyses on the agreement of the posterior distributions of~$\Nmultipole$ with~$\Nphase$. \textit{Left:}~One-dimensional, normalized posterior distributions of~$\Nmultipole$~(\textit{top}) and~$\Nphase$~(\textit{bottom}) for the $\Lambda\mathrm{CDM} + \Neff^{\delta X}$~model, with $X = \ell, \phi$ and~$\Neff$ fixed to~3.044. The results are shown for the `P18~[\texttt{plik}]' likelihood in~\textcolor{tabblue}{blue}, `P21~[\texttt{CamSpec}]' in~\textcolor{tabcyan}{cyan} and `P21~[\texttt{HiLLiPoP}]' in~\textcolor{navy}{navy}, with dashed and solid lines representing~$\Nmultipole$ and~$\Nphase$, respectively. \textit{Right:}~Two-dimensional joint constraints on~$\Neff$ and~$\Neff^{\delta X}$ for the same likelihood configurations. Dashed and solid contours correspond to the spectrum-~($X=\ell$) and perturbation-based~($X=\phi$) methods, respectively. The consistency of the posteriors of~$\Nmultipole$ with~$\Nphase$ in the updated Planck~2021 analyses underscores the consistency between the two methodologies and the improved reliability of the PBT~constraints.}\vspace{-1pt}
	\label{fig:sbt_vs_pbt_P21_pr4}
\end{figure}

Our analyses show that the differences between the spectrum- and perturbation-based methods effectively vanish on the Planck~2021~data. This is illustrated in Fig.~\ref{fig:sbt_vs_pbt_P21_pr4},%
where we present a detailed comparison of the resulting posterior distributions of the phase shift measurement for the three different Planck likelihood configurations considered in this work. In particular, we display the one-dimensional posterior distributions of~$\Nmultipole$ and~$\Nphase$ for the $\Lambda\mathrm{CDM} + \Neff^{\delta X}$~model in the left panel and directly compare the results from the `P18~TTTEEE'~(`P18~[\texttt{plik}]' hereafter for clarity), `P21~[\texttt{Camspec}]' and `P21~[\texttt{HiLLiPoP}]'~likelihood configurations. In the right panel, we instead compare the two-dimensional joint constraints on~$\Neff^{\delta X}$ and~$\Neff$, where both parameters are allowed to vary. For reference, we also provide a summary of the constraints on~$\Nmultipole$ and~$\Nphase$ inferred in these analyses in Table~\ref{tab:PR4_analysis}.%
\begin{table}
	\centering
	\begin{tabular}{l c c c c}
			\toprule
															& \multicolumn{2}{c}{P21 [\texttt{CamSpec}]}													& \multicolumn{2}{c}{P21 [\texttt{HiLLiPoP}]}													\\
															  \cmidrule(lr){2-3}																			  \cmidrule(lr){4-5}
															& {$\Nmultipole$}								& {$\Nphase$}									& {$\Nmultipole$}								& {$\Nphase$}									\\
			\midrule[0.065em]
		$\Lambda\mathrm{CDM} + \Neff^{\delta X}$			& $2.41^{+0.28\, (+0.62)}_{-0.33\, (-0.61)}$	& $2.33^{+0.28\, (+0.62)}_{-0.33\, (-0.59)}$	& $2.44^{+0.31\, (+0.66)}_{-0.34\, (-0.64)}$	& $2.34^{+0.28\, (+0.64)}_{-0.33\, (-0.61)}$	\\[4pt]
		$\Lambda\mathrm{CDM} + \Neff + \Neff^{\delta X}$	& $2.41^{+0.29\, (+0.63)}_{-0.35\, (-0.62)}$	& $2.35^{+0.29\, (+0.64)}_{-0.34\, (-0.61)}$	& $2.47^{+0.28\, (+0.65)}_{-0.36\, (-0.63)}$	& $2.35^{+0.28\, (+0.64)}_{-0.34\, (-0.62)}$	\\[2pt]
			\bottomrule
	\end{tabular}
	\caption{Constraints on~$\Nmultipole$ and~$\Nphase$ at 68\%~(95\%)~c.l.\ using the updated Planck~2021~likelihoods, as described in Sec.~\ref{sec:analysis-forecasts}. The two rows show the results for the scenarios where~$\Neff$ is either fixed to its SM~value or allowed to vary. In all cases, the six baseline parameters of the standard $\Lambda$CDM~model are marginalized.}
	\label{tab:PR4_analysis}
\end{table}

Overall, these comparisons demonstrate the remarkable consistency between the two methodologies when using the updated Planck~2021~likelihoods, for both the `P21~\texttt{CamSpec}' and `P21~\texttt{HiLLiPoP}' configurations. Interestingly, we find that the central values of the posterior distributions of~$\Nphase$ remain effectively unaltered between Planck~2018 and 2021~data. On the other hand, the central values of~$\Nmultipole$ shift between the two analyses, converging to the $\Nphase$~results. This highlights the enhanced robustness of the perturbation-based method relative to the spectrum-based framework. In fact, while we did not investigate in detail which specific feature in the Planck~2018~dataset is responsible for the elevated $\Nmultipole$~constraints,\footnote{This difference could in particular be related to the elevated level of smoothing of the acoustic peaks and troughs in the Planck~2018 spectra compared to the expected level based on the lensing spectrum, which is known as the $A_L$~anomaly~\cite{Planck:2018vyg}.} the consistency in the posterior distributions of~$\Nphase$ between the 2018~and 2021~datasets underscores its lower sensitivity to systematic effects. This suggests that the perturbation-based method provides a more reliable approach for extracting the neutrino-induced phase shift, as it is inherently less sensitive to potential artifacts or systematic features in the observed CMB~power spectra.

In addition, as expected, the reduced noise levels in the Planck~2021~data release lead to a modest improvement in our constraints on the neutrino-induced phase shift, tightening by approximately~20\% and yielding more symmetric~(Gaussian) error bars. This sharper precision, combined with the agreement of the posterior distributions of~$\Nmultipole$ with those for~$\Nphase$, results in the consistency with the standard physical model, $\Neff = \Neff^{\delta X} = 3.044$, at the $2\sigma$~level. This can particularly be seen in the analyses of the $\Lambda\mathrm{CDM} + \Neff^{\delta X}$ model, with~$\Neff$ fixed at~3.044, using Planck~2021~data. In these cases, the posterior distribution for the phase-shift parameter deviates between~$1.6\sigma$ and~$2.2\sigma$ from the expected SM~value $\Neff=3.044$, which is indicated by the dotted vertical lines in the left panels of Fig.~\ref{fig:sbt_vs_pbt_P21_pr4}. While this discrepancy remains small, it is intriguing, especially in light of upcoming CMB~experiments which will substantially tighten the phase-shift constraints as we will see in the following. In this sense, if the central values of the posterior distributions of the phase shift remain unchanged, this mild discrepancy could grow to a statistically significant level, potentially hinting at neutrinos not being fully free-streaming in the early universe prior to recombination due to new physics in the neutrino sector that could be probed in this way with upcoming surveys. It is however also worth pointing out that this mild discrepancy is observed exclusively in the analysis with Planck~2021 data alone. While we do not explicitly perform an analysis combining Planck~2021 with ACT~and SPT~data, we can infer the behavior by examining the perturbation-based results for \mbox{Planck~2018 + ACT + SPT}, which instead show consistency with the standard physical model well within the $1\sigma$~level.

\subsection{Forecasts}
\label{sec:forecasts}

Over the next decade, several experiments are scheduled to observe the CMB~anisotropies and especially polarization at significantly higher sensitivity than the current surveys from which we analyzed data above. In the following, we will therefore forecast the capabilities of upcoming and future CMB~observations to constrain the neutrino-induced phase shift using the well-established Fisher methodology.\footnote{We explicitly checked the reliability of our Fisher forecasts with an MCMC~analysis of lensed and unlensed power spectra for our \mbox{CMB-S4}~configuration. We find that the expected constraints generally differ by less than~$O(10\%)$ between the two methods, with the relative difference being less than a few percent for~$\Neff$ and~$\Nphase$. In addition, we observe that the inferred posterior distributions are close to Gaussian as expected for the tighter constraints.} We specifically consider the Simons Observatory~(SO)~\cite{SimonsObservatory:2018koc}, \mbox{\mbox{CMB-S4}}~(S4)~\cite{Abazajian:2019eic, CMB-S4:2022ght} and a cosmic-variance-limited~(CVL) experiment.\medskip

We expect that these future observations will lead to significant improvements in our constraining power since we will measure higher multipole modes with larger signal-to-noise ratios, which is particularly relevant for~$\Neff$ and~$\Nphase$. In fact, the parameter~$\Neff$ derives its sensitivity predominantly from the damping tail which is why more signal-dominated modes at larger multipoles will improve the constraints. The parameter~$\Nphase$ is measured through the acoustic peaks and troughs which implies that its constraints also improve with additional high-$\ell$, signal-dominated modes, but with diminishing improvements at very large multipoles since the acoustic oscillations are intrinsically damped themselves.\medskip

For an overview of the employed standard Fisher matrix approach for CMB~observations, we refer to Appendix~\ref{app:forecasts} where we also provide a detailed description of the experimental specifications and their implementation. In short, we take the noise curves for the baseline configuration of~SO from~\cite{SimonsObservatory:2018koc, SimonsObservatory:2020noise} and for \mbox{CMB-S4} from~\cite{CMB-S4:inprep, Raghunathan:draft},\footnote{The \mbox{CMB-S4}~experiment is currently undergoing a redesign of its configuration, with all telescopes being deployed to the Atacama desert in Chile instead of two sites in Chile and at the South Pole. We restrict our forecasts to the S4-wide survey with two large-aperture telescopes which will likely remain as previously planned. On the other hand, we do not take into account the S4-deep survey which was previously scheduled to be located at the South-Pole site and will now also be observing from Chile with a large-aperture telescope designed to optimally delens the B-mode measurements~(or the B-mode survey itself on large scales). Restricting our forecasts to the S4-wide survey is a conservative choice since these additional S4-deep observations from Chile will add further constraining power on small angular scales.} and we use~40\% and 62\%~of the sky, respectively, given the corresponding survey footprints and modeling of galactic foregrounds. We additionally combine these future noise curves with Planck-like noise spectra from~\cite{Allison:2015qca}.\footnote{The Planck-like specifications from~\cite{Allison:2015qca} lead to forecasted constraints on~$\Neff$ which are close to the achieved sensitivity with Planck~2018~\cite{Baumann:2017gkg}. While we have seen in~\textsection\ref{sec:comparison-data} that the Planck~2021 dataset is more constraining, the impact on our forecasts for future experiments would be minor since they are dominated by the new datasets.} Finally, we also consider a conservative CVL~experiment, for which we assume vanishing experimental noise over three quarters of the sky~(instead of covering the full sky to account for the Milky Way obscuring part of the~CMB) to estimate the potential futuristic reach of CMB~observations. For all experimental configurations, we take multipoles $\ell \leq 5000$ into account.

Another important aspect in forecasting the science reach of future CMB~experiments is the impact of weak gravitational lensing on the primary anisotropies~(cf.~\cite{Lewis:2006fu} for a comprehensive review). The CMB~photons are deflected by the large-scale structure of the universe which redistributes power in the CMB~spectra and results in a smoothing of the acoustic peaks. This in turn reduces the precision with which certain parameters can be inferred, especially those that benefit from the sharpness of the peaks, such as the phase shift. To largely mitigate this, we can reconstruct the lensing potential either internally within a CMB experiment or with the aid of external datasets~\cite{Okamoto:2003zw, Smith:2010gu, Larsen:2016wpa, Green:2016cjr, Sehgal:2016eag, Carron:2017vfg, ACT:2020goa, Millea:2020iuw, Hotinli:2021umk, SPT-3G:2024atg} and effectively remove the lensing contribution to the temperature and polarization power spectra. In this work, we use the internal delensing procedure at the level of these spectra established in~\cite{Green:2016cjr} which has been implemented \mbox{in \texttt{CLASS\_delens}~\cite{Hotinli:2021umk}~(see also~\cite{Trendafilova:2023xtq})}.\footnote{We also note that while delensing introduces non-Gaussian covariance terms, we can neglect these subdominant contributions in this work, since only parameters that are sensitive to the lensing power spectrum will be affected~\cite{Green:2016cjr}.}\medskip

We present our forecasted constraints for the Simons Observatory, \mbox{CMB-S4} and the CVL~experiment in Table~\ref{tab:forecasts}.
\begin{table}
	\centering
	\begin{tabular}{l S[table-format=1.3] S[table-format=1.3] S[table-format=1.3] S[table-format=1.3] S[table-format=1.3] S[table-format=1.3]}
			\toprule
						& {$\Lambda\mathrm{CDM}+\Neff$}	& {$\Lambda\mathrm{CDM}+\Nphase$}	& \multicolumn{2}{c}{$\Lambda\mathrm{CDM}+\Neff+\Nphase$}	& \multicolumn{2}{c}{$\Lambda\mathrm{CDM}+\Neff+\Nphase+Y_p$}	\\
						  \cmidrule(lr){2-2}			  \cmidrule(lr){3-3}				  \cmidrule(lr){4-5}										  \cmidrule(lr){6-7}
						& {$\sigma(\Neff)$}				& {$\sigma(\Nphase)$}				& {$\sigma(\Neff)$}		& {$\sigma(\Nphase)$}				& {$\sigma(\Neff)$}		& {$\sigma(\Nphase)$}					\\
			\midrule[0.065em]
		SO				& 0.054							& 0.13								& 0.054					& 0.14								& 0.29					& 0.14									\\
		\mbox{CMB-S4}	& 0.030							& 0.078								& 0.031					& 0.080								& 0.20					& 0.080									\\
		CVL				& 0.012							& 0.046								& 0.012					& 0.048								& 0.094					& 0.048									\\
			\bottomrule
	\end{tabular}
	\caption{Forecasted $1\sigma$~constraints on~$\Neff$ and~$\Nphase$ for the future CMB~experiments Simons Observatory and \mbox{CMB-S4} within $\Lambda\mathrm{CDM} + \Neff$, $\Lambda\mathrm{CDM} + \Nphase$, $\Lambda\mathrm{CDM} + \Neff + \Nphase$ and $\Lambda\mathrm{CDM} + \Neff + \Nphase + Y_p$. The underlying power spectra are delensed, with the noise curves for the experiments being based on the SO~baseline configuration and the S4-wide survey with a large galactic mask, respectively. In the bottom row, we also include a cosmic-variance-limited experiment with $f_\mathrm{sky} = 0.75$ and $\ell \leq 5000$.}
	\label{tab:forecasts}
\end{table}
Having established that SBT~and~PBT lead to consistent results on current data, which we also checked for lensed and unlensed spectra for~\mbox{CMB-S4}, we will focus on~PBT for simplicity. We see that the future surveys are projected to improve the constraints on~$\Nphase$ by factors of about approximately two and four for~SO and~\mbox{CMB-S4}, respectively, with respect to current data~(see Tables~\ref{tab:N_deltaphi_analysis} and~\ref{tab:PR4_analysis}), and by roughly~1.7 when going from~SO to~\mbox{CMB-S4}. This can be compared to the larger improvements by factors of roughly~three, five and~1.8, respectively, for constraints on~$\Neff$. We can understand this difference between~$\Neff$ and~$\Nphase$ from the fact that~$\Neff$ derives most of its constraining power from the background impact on the damping tail while the additional information on~$\Nphase$ is gradually diminished for larger multipoles due to the damping of the acoustic oscillations themselves. As with the current data, we observe that the constraints on~$\Neff$ and~$\Nphase$ are effectively uncorrelated. Similarly, the parameters~$\Nphase$ and~$Y_p$ are uncorrelated, but marginalizing over both of these parameters significantly weakens the constraints on~$\Neff$. The reason for this is that~$Y_p$ and~$\Nphase$ effectively capture two of the constraining signatures~(the damping tail and the phase shift, respectively) that would otherwise help to constrain~$\Neff$.

Looking towards the more distant future, our CVL~forecasts suggest that the constraints on the phase shift may improve by another factor of about two relative to those expected from~\mbox{CMB-S4}. By comparison, the constraints on~$\Neff$ are projected to tighten by a factor of three with the same experimental specifications. Similar to what we observed for~SO and~\mbox{CMB-S4}, this difference arises because the sensitivity to~$\Neff$ is predominantly driven by the damping tail, whereas the constraints on~$\Nphase$ rely on the acoustic oscillations. Consequently, the relative gain in constraining power for~$\Nphase$ is inherently more limited since the additional signal to noise will come from high multipoles in future experiments.\medskip

Finally, we also note that it would be interesting to combine these CMB~constraints with those from large-scale structure surveys in a joint analysis, especially from future galaxy or line-intensity-mapping observations~(cf.~\cite{CosmicVisions21cm:2018rfq, PUMA:2019jwd, Karkare:2022bai, DESI:2022lza, Schlegel:2022vrv}). While the LSS~forecasts from~\cite{Baumann:2017gkg} suggest that the CMB~temperature and polarization datasets discussed in this paper will dominate the sensitivity, incorporating LSS~information should still improve the respective constraints. Such a joint inference could follow the conventional approach of including constraints on the baryon-acoustic-oscillation parameters in a CMB~analysis. Alternatively, our newly developed perturbation-based approach would allow to model the CMB~and LSS~observables together. Either way, a joint analysis would leverage all cosmological datasets that contain the acoustic oscillations from the early universe with its neutrino-induced phase-shift signal.

\section{Conclusions and Outlook}
\label{sec:conclusions}

Standard Model neutrinos are expected to have been free-streaming through the cosmos ever since they decoupled from the primordial plasma at a temperature of about~\SI{1}{\mega\electronvolt}. This particular property of the cosmic neutrino background~(and other free-streaming relativistic species beyond the Standard Model of particle physics) especially imprints a characteristic phase shift in the acoustic oscillations as seen in the cosmic microwave background anisotropies and the large-scale structure of the universe. In this work, we establish and apply robust template-based analyses to infer the size of this effect from current and future CMB~datasets.\medskip

We present two complementary approaches for detecting the neutrino-induced phase shift. The first method, which we refer to as the spectrum-based approach, builds on the work of~\cite{Follin:2015hya} for the temperature data and parameterizes this effect using a multipole-dependent template that shifts the entire CMB~temperature and polarization power spectra according to the effective number of relativistic multipole-shifting species~$\Nmultipole$. The second, novel method, which we refer to as the perturbation-based approach, directly measures the phase shift in the photon-baryon perturbations where it is originally imprinted. Employing this wavenumber-dependent template, which shifts the CMB~perturbations according to the effective number of relativistic phase-shifting species~$\Nphase$, therefore is a more fundamental way to measure the phase-shift signature of free-streaming radiation in observed data.

Using these pipelines, we find overwhelming evidence for a non-zero phase shift in current observations of the CMB~temperature and polarization power spectra from the Planck~satellite, the Atacama Cosmology Telescope and the South Pole Telescope. With Planck~2018 data alone, the absence of a phase shift is excluded at a significance of~$10\sigma$, which increases to~$14\sigma$ when the additional ground-based datasets with higher-multipole information are included. In these analyses, the data are consistent with the Standard Model prediction of three free-streaming neutrinos, $\Nphase = \Nmultipole = \Neff^\mathrm{SM} = 3.044$, at roughly the $1\sigma$~level. We also analyze the Planck~NPIPE~2021 datasets which exclude a vanishing phase shift at a statistical significance of more than~$11\sigma$ by themselves and show consistency with three free-streaming neutrinos within approximately~$2\sigma$ confidence due to a shift in the mean values and smaller uncertainties. Our analysis further corroborates the critical role of polarization data in tightening the phase shift constraints since the sharper E-mode peaks and troughs enhance its detectability. In addition, we again confirm the robustness of the phase-shift signature as a probe of free-streaming radiation since~$\Nmultipole$ and~$\Nphase$ are minimally correlated with most of the other parameters in our considered cosmologies. The phase-shift parameters in particular have little correlation with~$\Neff$ and the Helium fraction~$Y_p$ which are instead degenerate with each other. The notable exception is the angular size of the sound horizon~$\theta_s$ which exhibits a strong anti-correlation since it effectively measures the frequency of the acoustic oscillations.

The consistency between the spectrum- and perturbation-based methods, validated first through mock-data analyses and subsequently confirmed on observed datasets, underscores the reliability of both approaches. While we identified a minor discrepancy between the two templates applied to the Planck~2018 dataset, with the spectrum-based approach yielding somewhat larger phase-shift values, this difference essentially vanishes when using the updated Planck~NPIPE~2021 data. This might point to artifacts or systematics in the earlier dataset as has also been seen in other analyses. Furthermore, we find that the constraints inferred using the perturbation-based template remain remarkably stable across both 2018~and 2021~datasets, demonstrating the improved robustness of our newly developed method.

Looking to the future, we forecasted the capabilities of upcoming and future CMB experiments. We in particular find that the Simons Observatory and~\mbox{CMB-S4} will significantly tighten the constraints on the neutrino-induced phase shift, with a 50\%~improvement over current observations with the former and an additional 40\%~improvement with~\mbox{CMB-S4} since we will measure higher multipoles at larger signal-to-noise ratios. These improvements also stem from the ability to delens the observed spectra at a high efficiency resulting in significantly sharper acoustic peaks and, therefore, phase-shift measurements. These advancements will enable unprecedented precision and accuracy in measuring this signature of free-streaming relativistic species, considerably better than the general constraints on the radiation density today. Our analysis pipeline will therefore also enable a robust detection of any potential deviations from the Standard Model expectation of three free-streaming neutrinos in a signature-driven and model-agnostic way. This could be further enhanced through a joint analysis of CMB~and large-scale-structure datasets based on the work in this paper and~\cite{Baumann:2017gkg, Baumann:2019keh, Whitford:2024ecj}. With the enhanced sensitivity of these next-generation experiments, the currently small potential deviations from the Standard Model expectation of three free-streaming neutrinos could achieve sufficient statistical significance to reveal insights into how neutrinos and other light relics propagated in the early universe. This could for example point to specific non-standard neutrino interactions, including flavor-dependent self-interactions or direct couplings to a dark sector~\cite{Jeong:2013eza, Chacko:2016kgg, Buen-Abad:2017gxg, Green:2021gdc, Berryman:2022hds, Adshead:2022ovo, Buen-Abad:2024tlb, Kaplan:2024ydw}. More generally, these future model-agnostic phase-shift measurements would then motivate exploring these and other BSM~scenarios in more detail.\medskip

To conclude, our results further establish the phase shift as a robust and sensitive probe of free-streaming relativistic species. The methods developed in this study, particularly the perturbation-based template, offer a general and powerful framework not only for constraining Standard Model neutrinos and free-streaming dark radiation, but also for exploring new physics more generally. These insights further highlight the extremely valuable and rewarding interplay between high-precision observational data, deep theoretical insights and statistical advances, paving the way for a refined understanding of the early universe and transformative discoveries with upcoming cosmological surveys.

\vskip20pt
\paragraph{Acknowledgments}
We are grateful to Kimberly Boddy, Raphael Flauger, Martina Gerbino, Subhajit Ghosh, Daniel Green, Gilbert Holder, Lloyd Knox, Massimiliano Lattanzi, Marilena LoVerde, Joel Meyers, Julian Muñoz, Vivian Poulin, Srinivasan Raghunathan, Cynthia Trendafilova and Yuhsin Tsai for useful discussions. We also thank Thejs Brinckmann, Nils Schöneberg and Vivian Poulin for help with the implementation of the Planck~2021 likelihoods in \texttt{MontePython}. K.\,F.~and G.\,M.~are grateful for support from the Jeff \& Gail Kodosky Endowed Chair in Physics held by~K.\,F.\ at the University of Texas at Austin. K.\,F.~and G.\,M.~also acknowledge funding from the US~Department of Energy under Grant~\mbox{DE-SC-0022021}. G.\,M. is also supported by the Continuing Fellowship of the Graduate School of the College of Natural Sciences at the University of Texas at Austin. K.\,F.~and B.\,W.~are supported by the Swedish Research Council under Contract No.~\mbox{638-2013-8993}. Nordita is supported in part by NordForsk. G.\,M.~would like to thank the Oscar Klein Centre at Stockholm University for hospitality. B.\,W.~is grateful to INFN~Ferrara, Scuola Normale Superiore di Pisa, University of California San Diego and University of Illinois Urbana-Champaign for their hospitality. This research is partly based on observations obtained with the Planck satellite~(\href{http://www.esa.int/Planck}{http:/\!/www.esa.int/Planck}), an ESA~science mission with instruments and contributions directly funded by ESA~Member States, NASA and Canada. This is not an official paper of the Simons Observatory Collaboration. Parts of this work were performed using computing resources provided by the Texas Advanced Computing Center~(TACC) at The University of Texas at Austin, and by the National Academic Infrastructure for Supercomputing in Sweden~(NAISS) under Projects~\mbox{2023/3-21} and~\mbox{2023/6-297}, which is partially funded by the Swedish Research Council through Grant~\mbox{2022-06725}. We acknowledge the use of \texttt{CLASS}~\cite{Blas:2011rf}, \texttt{CLASS\_delens}~\cite{Hotinli:2021umk}, \texttt{IPython}~\cite{Perez:2007ipy} and \texttt{MontePython}~\cite{Audren:2012wb, Brinckmann:2018cvx}, and the Python packages \texttt{Matplotlib}~\cite{Hunter:2007mat}, \texttt{NumPy}~\cite{Harris:2020xlr} and~\texttt{SciPy}~\cite{Virtanen:2019joe}.

\clearpage
\appendix
\section{Further Validation of the Analysis Pipeline}
\label{app:validation}

In this appendix, we provide additional details on a few aspects of the implementation and validation of our analysis pipeline that were not covered in the main text. These include considerations specific to the perturbation-based template~(\textsection\ref{app:pbt-details}), the impact of marginalizing over the parameters of the phase-shift template~(\textsection\ref{app:pbt-marginalization}), and joint constraints on~$\Neff$ and~$\Neff^{\delta X}$, $X = \ell, \phi$, inferred from mock data~(\textsection\ref{app:joint-constraints}).

\subsection{Additional Details on the Perturbation-Based Template}
\label{app:pbt-details}

When we implemented the perturbation-based template for the neutrino-induced phase shift in~\texttt{CLASS}, as described in~\textsection\ref{sec:sources_to_Cells}, we made some simplifying assumptions to ensure computational feasibility without compromising on accuracy. In the following, we briefly discuss two minor approximations that we did not discussed in detail in~\textsection\ref{sec:sources_to_Cells}.

\subsubsection*{Redshift-Dependence of the Phase Shift}

The perturbation-based template~$f_\phi(k, z)$ of~\eqref{eq:f_phi_def} is not only a function of wavenumber~$k$, but also of redshift~$z$. When including the relative phase shift~$\Delta\phi$ of~\eqref{eq:delta_phi_def2} in the relevant perturbations as described in~\eqref{eq:deltaphi_data}, we use the pre-computed template in the redshift range $500 \leq z \leq 2100$, as displayed in Fig.~\ref{fig:template_PBT}. For earlier and later times, we approximate~$f_\phi$ by the derived template at $z = 2100$, $f_\phi(k, z > 2100) = f_\phi(k, z = 2100)$, and a vanishing shift, $f_\phi(k, z < 500) = 0$, respectively. These choices are motivated by our theoretical expectation that the phase shift becomes constant~(negligible) deep in radiation~(matter) domination~\cite{Baumann:2015rya}.

We note that the specific choice of redshift range outside of which we apply the described approximation is somewhat arbitrary and simply reflects practical considerations that ensure being compatible with the standard precision settings of~\texttt{CLASS}. Having said that, we find that the induced multipole shift in the CMB~spectra is effectively determined by the evolution of the perturbations within $700 \lesssim z \lesssim 1300$ due to the sharply peaked nature of the visibility function~$g_\gamma(\eta)$ around recombination. In fact, we numerically confirmed that our choices for redshifts $z \notin [500, 2100]$ have negligible impact on the resulting multipole shift in the CMB~spectra and, consequently, on the measured constraints on~$\Nphase$.

\subsubsection*{Phase Shift in the Baryon Velocity Divergence}

We approximate the phase shift in the baryon velocity divergence~$\theta_b/k$ and its derivative~$\theta_b^\prime/k^2$ at late times. Specifically, we assume that the shift induced in these quantities is identical to that in the Sachs-Wolfe term,~$\Theta_0+\Psi$, and the polarization field~$\Pi$ across all redshifts. This is correct at early times when photons and baryons are tightly coupled. On the other hand, this is not strictly true at late times after the baryons have completely decoupled from the photons long after recombination. From then onward, the phase shift in~$\theta_b$ no longer follows the same time and mode dependence as in~$\Theta_0 + \Psi$ and~$\Pi$. This deviation however only becomes appreciable for $z \lesssim 900$. As with our other approximations, we have numerically verified that this treatment has a negligible impact on the CMB~spectra, thereby also justifying this simplifying assumption.

\subsection{Marginalization over the Template Parameters}
\label{app:pbt-marginalization}

\begin{figure}[b]
	\centering
	\includegraphics{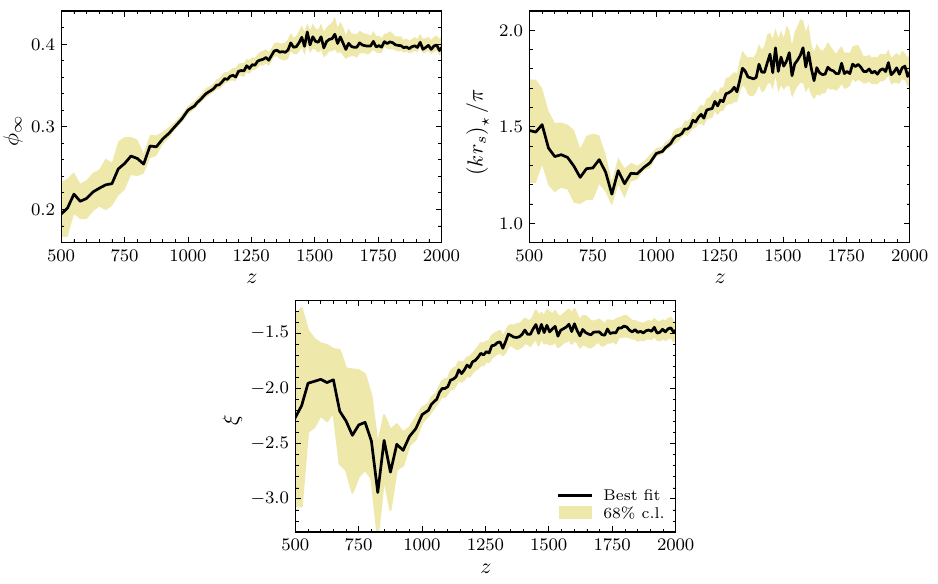}
	\caption{Redshift evolution of the fitting parameters~$\phi_\infty$, $(k r_s)_\star$ and~$\xi$, which characterize the perturbation-based template~$f_\phi(k, z)$ defined in~\eqref{eq:f_phi_def}. We show the best-fit values of these parameters along with their corresponding $1\sigma$~uncertainties as a function of redshift~$z$.}
	\label{fig:PBT_scatter_vs_z}
\end{figure}

The implementation of the multipole and $k$-mode shifts in our spectrum- and perturbation-based analysis pipelines relies on the phase-shift templates~\eqref{eq:f_ell_def} and~\eqref{eq:f_phi_def} with the numerically derived best-fit values for their parameters. In this setup, the shape of the templates is fixed by these best-fit parameters when constraining the phase shift in CMB~data. Our analysis therefore explicitly infers only the size of the phase shift, parameterized by~$\Nmultipole$ for~SBT and~$\Nphase$ for~PBT. As can however be seen in Figures~\ref{fig:template_SBT} and~\ref{fig:PBT_scatter_vs_z},%
the fitting parameters of the templates~$f_\ell$ and~$f_\phi$ have non-negligible uncertainties. It is therefore important to assess how these uncertainties may propagate into the phase-shift measurements and to determine whether fixing these parameters might lead to overly optimistic constraints.

To test this, we conducted an additional spectrum-based analysis. We performed an MCMC~inference of the `P18~TTTEEE'~dataset within the $\Lambda\mathrm{CDM} + \Nmultipole$~model and additionally marginalized over the three template parameters~$\phi_\infty$, $\ell_\star$ and~$\xi$ using Gaussian priors based on their $1\sigma$~fitting errors. Our results indicate that both the inferred central value and the uncertainty on~$\Nmultipole$ remain effectively unchanged compared to the analysis presented in~\textsection\ref{sec:spectrum-analysis} where the template parameters were fixed. This confirms that ignoring the uncertainties in the template parameters has a negligible impact, fully justifying our approach. While it is computationally impractical to directly marginalizing over the PBT~fitting parameters since they depend on redshift, the reduced uncertainties inherent to this method~(see Fig.~\ref{fig:template_PBT} and the discussion in~\textsection\ref{sec:perturbation-template}) also ensure its robustness in light of these SBT-based results. Both of our approaches can consequently provide reliable and accurate constraints on the phase shift from observational data.

\subsection[Joint Mock-Data Constraints on \texorpdfstring{$\Neff$}{Neff} and \texorpdfstring{$\Neff^{\delta\hskip-0.5pt X}$}{Neff-deltaX}]{Joint Mock-Data Constraints on $\mathbf{N}_\mathbf{eff}$ and $\mathbf{N}_\mathbf{eff}^{\boldsymbol{\delta}\mathbf{X}}$}
\label{app:joint-constraints}

\begin{figure}
	\centering
	\includegraphics{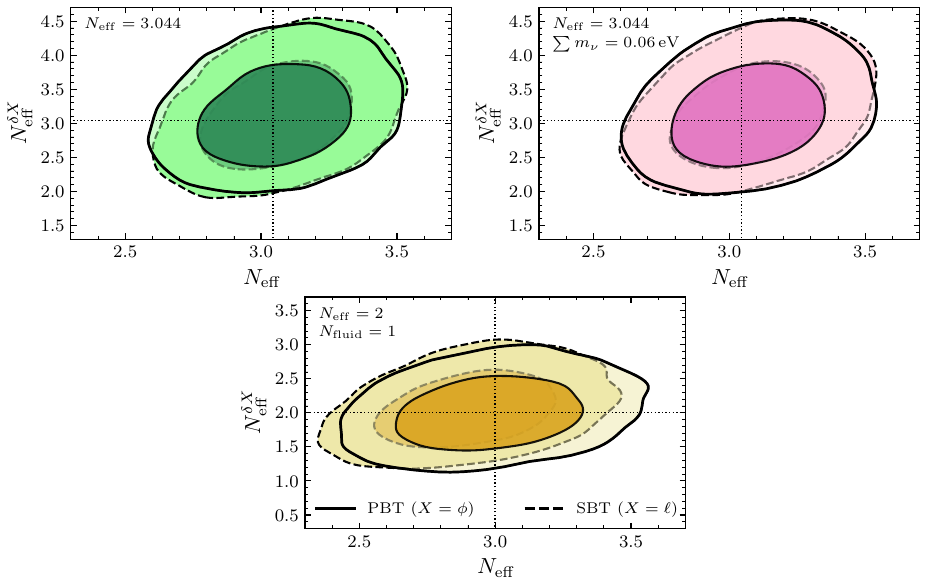}
	\caption{Two-dimensional joint constraints on~$\Neff$ and~$\Neff^{\delta X}$, $X = \ell,\phi$, for the three mock datasets described in~\textsection\ref{sec:validation_comparison}. The top panels show the posteriors for the scenarios with three massless free-streaming neutrinos with $\Neff=3.044$~(\textit{left}, \textcolor{tabgreen}{green}), and for three massive free-streaming neutrinos with $\Neff = 3.044$ and $\sum m_\nu =\SI{0.06}{\electronvolt}$~(\textit{right}, \textcolor{tabpink}{pink}). The bottom panel displays the results for the case with two free-streaming species, $\Neff = 2$, and one fluid-like specie, $N_\mathrm{fluid} = 1$~(\textcolor{goldenrod}{gold}). Solid and dashed contours represent the perturbation-~($\Nphase$) and spectrum-based~($\Nmultipole$) methods, respectively, and the dotted lines indicate the fiducial values of the mock datasets.}
	\label{fig:mock_planck_2d}
\end{figure}

We discussed the analyses of three physical mock datasets to validate the perturbation- and spectrum-based methods in~\textsection\ref{sec:validation_comparison}. We in particular focused on the relevant scenarios with massless and massive neutrinos, and with free-streaming and fluid-like relativistic species. To complement the one-dimensional posterior distributions shown in Fig.~\ref{fig:mock_planck}, we present the corresponding two-dimensional joint constraints on~$\Neff$ and~$\Neff^{\delta X}$, $X = \ell, \phi$, in Fig.~\ref{fig:mock_planck_2d}.%
Overall, this figure further emphasizes the robustness of our pipeline and demonstrates the consistency between the spectrum- and perturbation-based methods in accurately recovering the expected phase shifts.

\clearpage
\section{Comprehensive Analysis Results}
\label{app:tables}

In this appendix, we collect the complete set of constraints on all of the cosmological parameters and models with~$\Neff^{\delta X}$, $X = \ell, \phi$, analyzed in this work. Tables~\ref{tab:sbt_all}%
\begin{table}[h]
	\centering
	\subfloat[$\Lambda\mathrm{CDM} + \Nmultipole$]{
		\sisetup{group-digits=false}
		\begin{tabular}{l l l l l}
				\toprule
			Parameter			& P18 TT					& P18 TTTEEE				& P18 + ACT + SPT			& P21 [\texttt{HiLLiPoP}]	\\
				\midrule[0.065em]
			$100\,\omega_b$		& $2.216 \pm 0.022$			& $2.235 \pm 0.015$			& $2.233 \pm 0.012$			& $2.223 \pm 0.013$			\\
			$\omega_c$			& $0.1202 \pm 0.0021$		& $0.1201 \pm 0.0014$		& $0.1197 \pm 0.0012$		& $0.1188\pm 0.0012$		\\
			$100\,\theta_s$		& $1.0425 \pm 0.0020$		& $1.0432 \pm 0.0015$		& $1.0426 \pm 0.0011$		& $1.0440 \pm 0.0012$		\\
			$\ln(\num{e10}\As)$	& $3.041 \pm 0.017$			& $3.046 \pm 0.016$			& $3.047 \pm 0.014$			& $3.043 \pm 0.015$			\\
			$\ns$				& $0.9645 \pm 0.0061$		& $0.9664 \pm 0.0047$		& $0.9694 \pm 0.0040$		& $0.9699 \pm 0.0043$		\\
			$\tau$				& $0.0524 \pm 0.0080$		& $0.0543 \pm 0.0078$		& $0.0523 \pm 0.0067$		& $0.0585 \pm 0.0065$		\\
			$\Nmultipole$		& $2.86^{+0.54}_{-0.73}$	& $2.67^{+0.39}_{-0.43}$	& $2.87^{+0.31}_{-0.35}$	& $2.44^{+0.31}_{-0.34}$	\\
				\bottomrule
		\end{tabular}
	}\\[4pt]
	\subfloat[$\Lambda\mathrm{CDM} + \Nmultipole + \Neff$]{
		\sisetup{group-digits=false}
		\begin{tabular}{l l l l l}
				\toprule
			Parameter			& P18 TT					& P18 TTTEEE				& P18 + ACT + SPT			& P21 [\texttt{HiLLiPoP}]	\\
				\midrule[0.065em]
			$100\,\omega_b$		& $2.213 \pm 0.032$			& $2.226 \pm 0.022$			& $2.209 \pm 0.019$			& $2.234 \pm 0.020$			\\
			$\omega_c$			& $0.1196 \pm 0.0040$		& $0.1185 \pm 0.0032$		& $0.1162 \pm 0.0025$		& $0.1208 \pm 0.0030$		\\
			$100\,\theta_s$		& $1.0426 \pm 0.0021$		& $1.0432 \pm 0.0015$		& $1.0430 \pm 0.0011$		& $1.0440 \pm 0.0012$		\\
			$\ln(\num{e10}\As)$	& $3.038 \pm 0.021$			& $3.040 \pm 0.020$			& $3.033 \pm 0.016$			& $3.050 \pm 0.017$			\\
			$\ns$				& $0.9625 \pm 0.0136$		& $0.9618 \pm 0.0093$		& $0.9580 \pm 0.0080$		& $0.9749 \pm 0.0086$		\\
			$\tau$				& $0.0519 \pm 0.0083$		& $0.0537 \pm 0.0079$		& $0.0506 \pm 0.0063$		& $0.0596 \pm 0.0064$		\\
			$\Neff$				& $3.00^{+0.28}_{-0.30}$	& $2.93^{+0.20}_{-0.20}$	& $2.78^{+0.16}_{-0.16}$	& $3.18^{+0.19}_{-0.20}$	\\
			$\Nmultipole$		& $2.82^{+0.57}_{-0.82}$	& $2.64^{+0.38}_{-0.45}$	& $2.68^{+0.29}_{-0.35}$	& $2.47^{+0.28}_{-0.36}$	\\
				\bottomrule
		\end{tabular}
	}\\[4pt]
	\subfloat[$\Lambda\mathrm{CDM} + \Nmultipole + \Neff + Y_p$]{
		\sisetup{group-digits=false}
		\begin{tabular}{l l l l l}
				\toprule
			Parameter			& P18 TT					& P18 TTTEEE				& P18 + ACT + SPT			& P21 [\texttt{HiLLiPoP}]	\\
				\midrule[0.065em]
			$100\,\omega_b$		& $2.213 \pm 0.032$			& $2.227 \pm 0.023$			& $2.210 \pm 0.019$			& $2.236 \pm 0.020$			\\
			$\omega_c$			& $0.1270 \pm 0.0152$		& $0.1228 \pm 0.0071$		& $0.1215 \pm 0.0067$		& $0.1268 \pm 0.0068$		\\
			$100\,\theta_s$		& $1.0429 \pm 0.0023$		& $1.0433 \pm 0.0015$		& $1.0430 \pm 0.0011$		& $1.0440 \pm 0.0013$		\\
			$\ln(\num{e10}\As)$	& $3.046 \pm 0.030$			& $3.046 \pm 0.021$			& $3.041 \pm 0.018$			& $3.057 \pm 0.019$			\\
			$\ns$				& $0.9663 \pm 0.0163$		& $0.9645 \pm 0.0099$		& $0.9608 \pm 0.0087$		& $0.9787 \pm 0.0090$		\\
			$\tau$				& $0.0523 \pm 0.0084$		& $0.0532 \pm 0.0080$		& $0.0505 \pm 0.0062$		& $0.0590 \pm 0.0066$		\\
			$\Neff$				& $3.50^{+0.85}_{-1.24}$	& $3.22^{+0.42}_{-0.52}$	& $3.13^{+0.38}_{-0.49}$	& $3.59^{+0.39}_{-0.50}$	\\
			$\Nmultipole$		& $2.76^{+0.59}_{-0.86}$	& $2.62^{+0.36}_{-0.46}$	& $2.68^{+0.31}_{-0.36}$	& $2.47^{+0.32}_{-0.35}$	\\
			$Y_p$				& $0.217 \pm 0.067$			& $0.229 \pm 0.026$			& $0.223 \pm 0.024$			& $0.228 \pm 0.022$			\\
				\bottomrule
		\end{tabular}
	}
	\caption{Mean values and $1\sigma$~uncertainties for all the parameters of the extended cosmological models analyzed in this work using the spectrum-based template, $\Lambda\mathrm{CDM} + \Nmultipole$, $\Lambda\mathrm{CDM} + \Nmultipole + \Neff$ and $\Lambda\mathrm{CDM} + \Nmultipole + \Neff + Y_p$. The results are presented for the main dataset combinations described in Sec.~\ref{sec:analysis-forecasts}.}
	\label{tab:sbt_all}
\end{table}
and~\ref{tab:pbt_all}
\begin{table}
	\centering
	\subfloat[$\Lambda\mathrm{CDM} + \Nphase$]{
		\sisetup{group-digits=false}
		\begin{tabular}{l l l l l}
				\toprule
			Parameter			& P18 TT					& P18 TTTEEE				& P18 + ACT + SPT			& P21 [\texttt{HiLLiPoP}]	\\
				\midrule[0.065em]
			$100\,\omega_b$		& $2.2201 \pm 0.0236$		& $2.2401 \pm 0.0159$		& $2.2352 \pm 0.0121$		& $2.229 \pm 0.013$			\\
			$\omega_c$			& $0.1203 \pm 0.0022$		& $0.1200 \pm 0.0014$		& $0.1197 \pm 0.0012$		& $0.01187 \pm 0.0012$		\\
			$100\,\theta_s$		& $1.0432 \pm 0.0022$		& $1.0440 \pm 0.0014$		& $1.0431 \pm 0.0011$		& $1.0443 \pm 0.0012$		\\
			$\ln(\num{e10}\As)$ & $3.0408 \pm 0.0163$		& $3.0450 \pm 0.0165$		& $3.0470 \pm 0.0136$		& $3.040 \pm 0.014$			\\
			$\ns$				& $0.9642 \pm 0.0060$		& $0.9664 \pm 0.0046$		& $0.9694 \pm 0.0039$		& $0.9692 \pm 0.0043$		\\
			$\tau$				& $0.0524 \pm 0.0080$		& $0.0543 \pm 0.0079$		& $0.0522 \pm 0.0067$		& $0.0580 \pm 0.0063$		\\
			$\Nphase$			& $2.63^{+0.53}_{-0.83}$	& $2.41^{+0.34}_{-0.42}$	& $2.71^{+0.30}_{-0.34}$	& $2.34^{+0.28}_{-0.33}$	\\
				\bottomrule
		\end{tabular}
	}\\[4pt]
	\subfloat[$\Lambda\mathrm{CDM} + \Nphase + \Neff$]{
		\sisetup{group-digits=false}
		\begin{tabular}{l l l l l}
				\toprule
			Parameter			& P18 TT					& P18 TTTEEE				& P18 + ACT + SPT			& P21 [\texttt{HiLLiPoP}]	\\
				\midrule[0.065em]
			$100\,\omega_b$		& $2.2156 \pm 0.0341$		& $2.2358 \pm 0.0239$		& $2.2127 \pm 0.0195$		& $2.249 \pm 0.0217$		\\
			$\omega_c$			& $0.1198 \pm 0.0041$		& $0.1194 \pm 0.0033$		& $0.1163 \pm 0.0025$		& $0.1221 \pm 0.0032$		\\
			$100\,\theta_s$		& $1.0433 \pm 0.0022$		& $1.0440 \pm 0.0014$		& $1.0431 \pm 0.0011$		& $1.0444 \pm 0.0017$		\\
			$\ln(\num{e10}\As)$	& $3.0383 \pm 0.0216$		& $3.0430 \pm 0.0194$		& $3.0328 \pm 0.0157$		& $3.0505 \pm 0.0171$		\\
			$\ns$				& $0.9619 \pm 0.0139$		& $0.9646 \pm 0.0095$		& $0.9587 \pm 0.0081$		& $0.9783 \pm 0.0086$		\\
			$\tau$				& $0.0521 \pm 0.0084$		& $0.0541 \pm 0.0080$		& $0.0505 \pm 0.0063$		& $0.0591 \pm 0.0065$		\\
			$\Neff$				& $3.00^{+0.28}_{-0.30}$	& $3.00^{+0.21}_{-0.21}$	& $2.80^{+0.16}_{-0.16}$	& $3.29^{+0.20}_{-0.21}$	\\
			$\Nphase$			& $2.59^{+0.58}_{-0.81}$	& $2.41^{+0.33}_{-0.42}$	& $2.62^{+0.30}_{-0.35}$	& $2.35^{+0.28}_{-0.34}$	\\
				\bottomrule
		\end{tabular}
	}\\[4pt]
	\subfloat[$\Lambda\mathrm{CDM} + \Nphase + \Neff + Y_p$]{
		\sisetup{group-digits=false}
		\begin{tabular}{l l l l l}
				\toprule
			Parameter			& P18 TT					& P18 TTTEEE				& P18 + ACT + SPT			& P21 [\texttt{HiLLiPoP}]	\\
				\midrule[0.065em]
			$100\,\omega_b$		& $2.2211 \pm 0.0355$		& $2.2393 \pm 0.0247$		& $2.2169 \pm 0.0197$		& $2.2550 \pm 0.0212$		\\
			$\omega_c$			& $0.1268 \pm 0.0113$		& $0.1244 \pm 0.0071$		& $0.1228 \pm 0.0066$		& $0.1294 \pm 0.0069$		\\
			$100\,\theta_s$		& $1.0436 \pm 0.0023$		& $1.0440 \pm 0.0014$		& $1.0432 \pm 0.0011$		& $1.0444 \pm 0.0012$		\\
			$\ln(\num{e10}\As)$	& $3.0460 \pm 0.0257$		& $3.0492 \pm 0.0212$		& $3.0420 \pm 0.0176$		& $3.0599 \pm 0.0184$		\\
			$\ns$				& $0.9653 \pm 0.0150$		& $0.9671 \pm 0.0101$		& $0.9621 \pm 0.0087$		& $0.9822 \pm 0.0091$		\\
			$\tau$				& $0.0524 \pm 0.0084$		& $0.0539 \pm 0.0081$		& $0.0503 \pm 0.0062$		& $0.0592 \pm 0.0066$		\\
			$\Neff$				& $3.48^{+0.53}_{-1.04}$	& $3.34^{+0.37}_{-0.57}$	& $3.23^{+0.39}_{-0.49}$	& $3.78^{+0.45}_{-0.49}$	\\
			$\Nphase$			& $2.50^{+0.57}_{-0.83}$	& $2.42^{+0.35}_{-0.42}$	& $2.62^{+0.30}_{-0.35}$	& $2.36^{+0.28}_{-0.34}$	\\
			$Y_p$				& $0.217 \pm 0.046$			& $0.228 \pm 0.023$			& $0.220 \pm 0.023$			& $0.225\pm 0.022$			\\
				\bottomrule
		\end{tabular}
	}
	\caption{Same as Table~\ref{tab:sbt_all}, but for the perturbation-based analyses.}
	\label{tab:pbt_all}
\end{table}
summarize the results from our MCMC~analyses of the three main dataset combinations of current CMB~measurements involving Planck~2018 data described in~Sec.~\ref{sec:analysis-forecasts}, for the spectrum- and perturbation-based approaches, respectively. In addition, we include the constraints inferred from the updated Planck~\texttt{NPIPE} data as incorporated in the Planck~2021 \texttt{HiLLiPoP}~likelihood. On the other hand, we omit those from the Planck~2021 \texttt{CamSpec}~likelihood since the inferences lead to very similar posterior distributions. We clearly see how high-$\ell$ polarization data from Planck significantly improves the determination of various cosmological parameters, including the phase-shift parameters~$\Neff^{\delta X}$, which further tighten with the addition of smaller-scale~ACT and SPT~data. Of notable relevance is the addition of polarization information to the `P18~TT' analyses resulting in a significant improvement not only in the constraints on~$\Neff$ and~$\Nphase$, but particularly also on the primordial helium fraction~$Y_p$ since the polarization amplitude is sensitive to~$Y_p$ and not~$\Neff$, unlike the damping tail~\cite{Baumann:2015rya}. Finally, we observe that the Planck~2021 \texttt{HiLLiPoP}~analysis yields about 10--20\% smaller uncertainties on parameter constraints, relative to the P18-only analysis, consistent with the reduced noise levels of the \texttt{NPIPE}~data release. Moreover, the posteriors for~$\Neff$ shift by approximately~$1\sigma$ towards larger value while those for~$\Nmultipole$ and~$\Nphase$ remain effectively unchanged~(see~\textsection\ref{sec:comparison-data} for further discussion), resulting in a roughly $2\sigma$~deviation from the expected SM~value $\Neff^{\delta X} = \Neff = 3.044$.

\clearpage
\section{Details on the Forecasts}
\label{app:forecasts}

This appendix provides a more detailed overview of the Fisher forecasting methodology employed in this work to estimate the sensitivity of future CMB~observations to the phase shift. We also include further details on the employed experimental configurations and their implementation. In addition, we provide the forecasted constraints for all cosmological parameters and extended cosmologies that can be considered in the context of our work.\medskip

We employ a standard Fisher methodology to forecast the sensitivity of future CMB~experiments. This means that we calculate the covariance matrix of the Gaussian approximation to the posterior distribution around the fiducial cosmology and estimate the future constraints by applying the Cramér-Rao bound. As noted already in the main text, we checked the validity of this approximation through an MCMC~analysis of \mbox{CMB-S4}-like mock power spectra. In the context of CMB~experiments, the Fisher information matrix is given by
\begin{equation}
	F_{ij} = \sum_{a,b}\sum^{\ell_\mathrm{max}}_{\ell=\ell_\mathrm{min}}\frac{\partial C^a_\ell}{\partial \theta_i} \left[\mathbf{C}^{a b}_\ell\right]^{\!-1} \frac{\partial C^b_\ell}{\partial \theta_j}\, ,	\label{eq:Fisher}
\end{equation}
where~$\theta_i$ are the model parameters of interest, $C_\ell^a$~are the theoretical CMB~power spectra and $a, b \in \{TT, EE, TE\}$.\footnote{We ignore the lensing convergence since it does not impact the constraints on the radiation energy density and the neutrino-induced phase shift.}
The covariance matrix~$\mathbf{C}^{ab}_\ell$ for CMB~temperature, E-mode and cross-correlation spectra is defined as
\begin{equation}
	\mathbf{C}_{\ell} = \frac{2}{(2 \ell+1) f_\mathrm{sky}} \times \,
	\begingroup
	\scriptstyle \renewcommand*{\arraystretch}{1.2} \setlength{\arraycolsep}{6pt}
	\begin{pmatrix}
		\bigl(\tilde{C}_\ell^{TT}\bigr)^{\!2}	& \bigl(\tilde{C}_\ell^{TE}\bigr)^{\!2}		& \tilde{C}_\ell^{TT} \tilde{C}_\ell^{TE}																	\\
		\bigl(\tilde{C}_\ell^{TE}\bigr)^{\!2}	& \bigl(\tilde{C}_\ell^{EE}\bigr)^{\!2}		& \tilde{C}_\ell^{EE} \tilde{C}_\ell^{TE}																	\\
		\tilde{C}_\ell^{TT} \tilde{C}_\ell^{TE}	& \tilde{C}_\ell^{EE} \tilde{C}_\ell^{TE}	& \frac{1}{2} \bigl[\bigl(\tilde{C}_\ell^{TE}\bigr)^{\!2} + \tilde{C}_\ell^{TT} \tilde{C}_\ell^{EE}\bigr]
	\end{pmatrix}
	\endgroup ,
	\label{eq:Fisher_covariance}
\end{equation}
where $f_\mathrm{sky}$ is the observed fraction of the sky and $\tilde{C}_\ell^a \equiv C_{\ell}^a + N_\ell^a$ are the expected CMB~power spectra as observed, accounting for cosmic variance and the noise spectra~$N_\ell^a$ of a given experiment. We use the Fisher-matrix implementation that was previously employed in~\cite{Baumann:2017gkg, Beutler:2019ojk}, with spectra delensed using~\texttt{CLASS\_delens}.\medskip

The covariance matrix~\eqref{eq:Fisher_covariance} directly exhibits the two main limitations of any CMB~experiment: sample variance and observational noise. The former arises from the limited number of modes that we can observe at any given multipole~$\ell$ for the sky fraction~$f_\mathrm{sky} < 1$ accessible to a survey which is inherently smaller than the cosmic-variance limit that is set by~$C_\ell^a$. The latter is in particular dictated by the instrumental specifications of the CMB~detectors and telescopes, but may also include systematic and foreground effects. We also note that the employed internal delensing procedure does not rely on external datasets, but is limited by the noise levels of the respective CMB~experiment.

For the Simons Observatory, we take the noise curves from~\cite{SimonsObservatory:2018koc, SimonsObservatory:2020noise} based on a standard internal linear combination~(ILC) calculation using the baseline configuration of the experiment over~40\% of the sky~($f_\mathrm{sky} = 0.4$).\footnote{We note that the employed SO~noise curves from~\cite{SimonsObservatory:2020noise} were updated and slightly differ from those used in~\cite{SimonsObservatory:2018koc}.} The minimum and maximum multipoles are chosen to be $\ell_\mathrm{min} = 30$ and $\ell_\mathrm{max} = 5000$ since we impose a Gaussian prior on the optical depth~$\tau$ with width $\sigma(\tau) = 0.01$ instead of including low-$\ell$ polarization data. We additionally combine these noise curves with Planck-like noise spectra from~\cite{Allison:2015qca} using inverse-variance weighting when calculating the delensed spectra and performing our Fisher forecasts.

For~\mbox{CMB-S4}, we use the noise power spectra derived using the dark radiation anisotropy flowdown team~(DRAFT) tool~\cite{CMB-S4:inprep, Raghunathan:draft} which implements a standard ILC~approach including the galactic foreground contributions from dust and synchrotron, and extragalactic components from the cosmic infrared background, radio galaxies and the Sunyaev-Zel'dovich effects~(see also~\cite{Raghunathan:2023yfe}).\footnote{We obtained essentially the same forecasted constraints for~SO when applying the DRAFT~tool to the experimental specifications underlying the noise curves from~\cite{SimonsObservatory:2020noise}.} We restrict to the two large-aperture telescopes which have always been planned to be observing about 67\%~of the sky from the Chile site, with the experimental specifications taken from the preliminary baseline design report~\cite{CMB-S4:pbdr}. For our forecasts, we use an optimistic galactic mask over the S4~footprint, which is one of the choices discussed in~\cite{CMB-S4:inprep, Raghunathan:inprep} and leaves 62\%~of the sky~($f_\mathrm{sky} = 0.62$).\footnote{It has been shown that mismodeling extragalactic foregounds will not affect the sensitivity to neutrinos and other light relics~\cite{Raghunathan:2023yfe}, and we refer to~\cite{CMB-S4:inprep, Raghunathan:inprep} for a discussion and detailed study of the potential impacts of galactic foregrounds and their (mis)modeling.} As for~SO, the minimum and maximum multipoles are chosen to be $\ell_\mathrm{min} = 30$ and $\ell_\mathrm{max} = 5000$, and we also combine these noise curves with Planck-like noise spectra from~\cite{Allison:2015qca}.

For the conservative CVL~experiment considered in this work, we set the noise spectra to zero, $N_\ell^a \equiv 0$, and use $f_\mathrm{sky} = 0.75$ and $\ell \in [2, 5000]$. We note that this is actually a~(conservative) sample-variance-limited experiment since we only use 75\%~instead of 100\%~of the sky to account for the fact that our galaxy obscures part of the~CMB. In this context, we also mention that \mbox{CMB-HD}~\cite{CMB-HD:2022bsz}~proposes to reliably measure much larger multipoles which would further improve the sensitivity on the phase shift.\medskip

Finally, we summarize the results of our Fisher forecasts with delensed~TT, TE and~EE spectra for~SO, \mbox{CMB-S4} and the CVL~experiment in Table~\ref{tab:forecasts_all}%
\begin{table}
	\centering
	\subfloat[Simons Observatory]{
		\begin{tabular}{l S[table-format=1.1] S[table-format=2.1] S[table-format=2.1] S[table-format=1.1] S[table-format=1.3] S[table-format=1.3] S[table-format=1.5]}
				\toprule
			Model											& {$\num{e5}\,\omega_b$}	& {$\num{e4}\,\omega_c$}	& {$\num{e7}\,\theta_s$}	& {$\num{e3}\,\ns$}	& {$\Neff$}	& {$\Nphase$}	& {$Y_p$}	\\
				\midrule[0.065em]
			$\Lambda\mathrm{CDM}$							& 5.1						& 9.2						& 9.5						& 2.6				& {--} 		& {--}			& {--}		\\
			$\Lambda\mathrm{CDM} + \Neff$					& 8.1						& 11						& 12						& 4.4				& 0.054		& {--}			& {--}		\\
			$\Lambda\mathrm{CDM} + \Nphase$					& 5.1						& 9.2						& 30						& 2.6				& {--}		& 0.13			& {--}		\\
			$\Lambda\mathrm{CDM} + \Neff + \Nphase$			& 8.4						& 11						& 30						& 4.4				& 0.054		& 0.14			& {--}		\\
			$\Lambda\mathrm{CDM} + Y_p$						& 8.2						& 9.5						& 9.9						& 4.1				& {--}		& {--}			& 0.0037	\\
			$\Lambda\mathrm{CDM} + \Neff + Y_p$				& 8.3						& 21						& 27						& 4.4				& 0.13		& {--}			& 0.0071	\\
			$\Lambda\mathrm{CDM} + \Nphase + Y_p$			& 8.2						& 9.5						& 30						& 4.1				& {--}		& 0.14			& 0.0038	\\
			$\Lambda\mathrm{CDM} + \Neff + \Nphase + Y_p$	& 8.4						& 46						& 30						& 4.8				& 0.29		& 0.14			& 0.016		\\
				\bottomrule
		\end{tabular}
	}\\[4pt]
	\subfloat[CMB-S4]{
		\begin{tabular}{l S[table-format=1.1] S[table-format=2.1] S[table-format=2.1] S[table-format=1.1] S[table-format=1.3] S[table-format=1.3] S[table-format=1.5]}
				\toprule
			Model											& {$\num{e5}\,\omega_b$}	& {$\num{e4}\,\omega_c$}	& {$\num{e7}\,\theta_s$}	& {$\num{e3}\,\ns$}	& {$\Neff$}	& {$\Nphase$}	& {$Y_p$}	\\
				\midrule[0.065em]
			$\Lambda\mathrm{CDM}$							& 2.7						& 6.5						& 5.5						& 1.8				& {--}		& {--}			& {--}		\\
			$\Lambda\mathrm{CDM} + \Neff$					& 4.0						& 7.5						& 6.7						& 2.9				& 0.030		& {--} 			& {--}		\\
			$\Lambda\mathrm{CDM} + \Nphase$					& 2.7						& 6.5						& 16						& 1.8				& {--}		& 0.078			& {--}		\\
			$\Lambda\mathrm{CDM} + \Neff + \Nphase$			& 4.1						& 7.5						& 16						& 2.9				& 0.031		& 0.080			& {--}		\\
			$\Lambda\mathrm{CDM} + Y_p$						& 4.1						& 6.6						& 5.8						& 2.8				& {--}		& {--}			& 0.0021	\\
			$\Lambda\mathrm{CDM} + \Neff + Y_p$				& 4.1						& 13						& 14						& 2.9				& 0.076		& {--}			& 0.0044	\\
			$\Lambda\mathrm{CDM} + \Nphase + Y_p$			& 4.1						& 6.6						& 16						& 2.8				& {--}		& 0.079			& 0.0022	\\
			$\Lambda\mathrm{CDM} + \Neff + \Nphase + Y_p$	& 4.1						& 32						& 16						& 3.1				& 0.20		& 0.080			& 0.011		\\
				\bottomrule
		\end{tabular}
	}\\[4pt]
	\subfloat[Cosmic-variance-limited experiment]{
		\begin{tabular}{l S[table-format=1.1] S[table-format=2.1] S[table-format=2.1] S[table-format=1.1] S[table-format=1.3] S[table-format=1.3] S[table-format=1.5]}
				\toprule
			Model											& {$\num{e5}\,\omega_b$}	& {$\num{e4}\,\omega_c$}	& {$\num{e7}\,\theta_s$}	& {$\num{e3}\,\ns$}	& {$\Neff$}	& {$\Nphase$}	& {$Y_p$}	\\
				\midrule[0.065em]
			$\Lambda\mathrm{CDM}$							& 1.0						& 2.0						& 2.9						& 1.1				& {--}		& {--}			& {--}		\\
			$\Lambda\mathrm{CDM} + \Neff$					& 1.4						& 2.1						& 2.9						& 1.8				& 0.012		& {--} 			& {--}		\\
			$\Lambda\mathrm{CDM} + \Nphase$					& 1.1						& 2.1						& 7.2						& 1.1				& {--}		& 0.046			& {--}		\\
			$\Lambda\mathrm{CDM} + \Neff + \Nphase$			& 1.4						& 2.1						& 7.2						& 1.8				& 0.012		& 0.048			& {--}		\\
			$\Lambda\mathrm{CDM} + Y_p$						& 1.4						& 2.2						& 3.0						& 1.6				& {--}		& {--}			& 0.00080	\\
			$\Lambda\mathrm{CDM} + \Neff + Y_p$				& 1.4						& 5.2						& 5.5						& 1.8				& 0.042		& {--} 			& 0.0024	\\
			$\Lambda\mathrm{CDM} + \Nphase + Y_p$			& 1.4						& 2.2						& 7.2						& 1.7				& {--}		& 0.048			& 0.00082	\\
			$\Lambda\mathrm{CDM} + \Neff + \Nphase + Y_p$	& 1.4						& 11						& 7.7						& 2.0				& 0.094		& 0.048			& 0.0054	\\
				\bottomrule
		\end{tabular}
	}\vspace{-5pt}
	\caption{Summary of forecasted sensitivities for the future surveys and the extended cosmological models under consideration in this work. We show the $1\sigma$~constraints obtained in our Fisher forecasts based on delensed~TT, TE and EE~spectra for the baseline configuration of the Simons Observatory, the wide-field survey of~\mbox{CMB-S4} and our conservative CVL~experiment. The constraints on the scalar amplitude~$\As$ and the optical depth~$\tau$ \mbox{are mentioned in the main text.}}
	\label{tab:forecasts_all}
\end{table}
as a point of reference. This does not only include the parameters~$\Neff$, $\Nphase$ and~$Y_p$ which are the focus of our work, but also the $\Lambda$CDM~parameters~$\omega_b$, $\omega_c$, $\theta_s$ and~$n_s$. The reason why the scalar amplitude~$A_s$ and the optical depth~$\tau$ are not included in the table is that their forecasted constraints are essentially the same in all considered cosmologies: $\sigma(\ln(10^{10}\As)) \approx 0.020$ and $\sigma(\tau) = 0.010$ for~SO and \mbox{CMB-S4}, while these constraints improve to approximately~0.0031 and~0.0017 with our CVL~configuration.

Before adding a few additional comments on the relative behavior of the parameter constraints, we note that our approximations of neglecting the lensing power spectrum and the non-Gaussian covariance terms in our forecasts impact some of the projected values~(cf.~\cite{Green:2016cjr}). Nevertheless, these simplifications do not prevent us from identifying degeneracies and physical dependencies when comparing results across the various cosmological models. We additionally emphasize that~$\sigma(\Neff)$, $\sigma(\Nphase)$ and~$\sigma(Y_p)$ are not affected by this approximation~\cite{Green:2016cjr} since these are early-universe parameters and do not derive constraining power from the lensing power spectrum which is a late-time observable.

The degeneracies among the beyond-$\Lambda$CDM~parameters~$\Neff$, $\Nphase$ and~$Y_p$ have already been discussed in the main text and can further be observed in Table~\ref{tab:forecasts_all}, including the degraded constraints on~$Y_p$ when varying~$\Neff$. To summarize, while the phase-shift parameter~$\Nphase$ exhibits minimal correlation with both~$Y_p$ and~$\Neff$, the constraints on the latter are weakened when varying~$\Nphase$ and~$Y_p$. This occurs because~$\Neff$ effectively lacks the constraining power from the phase and damping scale, respectively. In addition, a clear degeneracy can be seen between the angular size of the sound horizon~$\theta_s$ and the phase parameter~$\Nphase$, with $\sigma(\theta_s)$ weakening by a factor of two to three depending on the experimental configuration. The origin of this degeneracy lies in the limited number of acoustic peaks and troughs available to simultaneously constrain the frequency and phase of the oscillations when inferring both~$\Nphase$ and~$\theta_s$. In fact, this degeneracy is reduced in more sensitive surveys since their higher signal-to-noise ratio at small scales allows for the measurement of additional peaks. To conclude, we reiterate that this comprehensive overview of future constraints once again highlights the robustness of the phase shift as a powerful probe of free-streaming neutrinos and other light relics.

\clearpage
\phantomsection
\addcontentsline{toc}{section}{References}
\bibliographystyle{utphys}
\bibliography{references}

\end{document}